\documentclass[twocolumn,prd,showpacs,preprintnumbers,amsmath,amssymb,amsfonts,nofootinbib]{revtex4}

\usepackage{graphicx}
\usepackage{amssymb}
\usepackage{psfrag}
\usepackage{ulem}
\usepackage{color}
\usepackage{hyperref}
\usepackage{natbib}
\definecolor{dr}{rgb}{1.00,0.0,.33}
\begin{document}

\title{Testing model independent modified gravity with future large
  scale surveys}

\author{Daniel B.~Thomas}
\email{daniel.b.thomas08@imperial.ac.uk}
\author{Carlo R.~Contaldi}
\affiliation{Theoretical Physics, Blackett Laboratory, Imperial
  College, London}
\date{\today}

\pacs{04.80.Cc,98.65.Cw,98.80.-k}

\begin{abstract}
  Model-independent parametrisations of modified gravity have
  attracted a lot of attention over the past few years and numerous
  combinations of experiments and observables have been suggested to
  constrain the parameters used in these models. Galaxy clusters have
  been mentioned, but not looked at as extensively in the literature
  as some other probes. Here we look at adding galaxy clusters into
  the mix of observables and examine how they could improve the
  constraints on the modified gravity parameters. In particular, we
  forecast the constraints from combining Planck satellite Cosmic
  Microwave Background (CMB) measurements and Sunyaev--Zeldovich (SZ)
  cluster catalogue with a DES-like Weak Lensing (WL) survey. We find
  that cluster counts significantly improve the constraints over those
  derived using CMB and WL. We then look at surveys further into the
  future, to see how much better it may be feasible to make the
  constraints.
\end{abstract}

\maketitle

\section{Introduction}
Einstein's General Relativity (GR) is one of the principal ingredients
of modern cosmology. Indeed, it could be argued that it was only with
the development of GR that cosmology really became a part of
physics. Nonetheless, it is our job as physicists to continue to test
even the most fundamental pillars of cosmology in order to refine, improve
and further justify our model of the universe. There are also
fundamental reasons for considering different theories of gravity: GR
is inconsistent with quantum mechanics and the search for a theory
of `Quantum Gravity' is one of the holy grails of modern physics.

It has proved to be difficult to test GR outside of the solar system,
particularly as the effects of a different theory of gravity could be
degenerate with behaviour induced by different constituents of the
universe. This is the case with current observations that suggest the
presence of some form of dark matter and dark energy. Dark matter is
required to explain galaxy rotation curves, galaxy lensing,
nucleosynthesis, acoustic oscillations in the CMB, and the growth of
structure in the Universe to name a few. The requirement for dark
energy is underpinned by observations of the background expansion rate
and large scale structure measurement.

The question has often been raised as to whether these effects could
be due to a modified gravity theory. Proposals for such a theory
include a number of $f(R)$ theories, the Dvali Gabadadze Porrati (DGP)
model \cite{dgp}, conformal gravity \cite{cfg}, Modified Newtonian
dynamics (MOND) \cite{mond} and its covariant, relativistic extensions
\cite{teves}, and Einstein Aether theories \cite{eatheories}, (see
also the recent review \cite{mgreview} for an exhaustive list of
candidates). It is possible, if GR is indeed the correct theory of
gravity, that this debate will only be settled by the
non-gravitational detection of the dark matter and/or dark
energy. However, since we do not know whether or not GR is indeed the
correct theory of gravity, it seems reasonable to consider which
observations could allow us to detect modified gravity.

More importantly, are there ways to test deviations from GR in a model
independent way? There are several advantages to a model independent
approach; some alternatives to GR do exist but there is no complete
theory of, for example, quantum gravity to draw on. Also, there are no
`stand-out' candidates that are universally considered to be strong
alternatives. Another advantage of a model independent approach is
that the results do not rely on model selection techniques such as
$\chi^2$ per degree of freedom or other ways of choosing between
competing theories. Thus, they also do not rely on us having {\sl the}
correct theory to hand. A result that is inconsistent with the GR based,
concordance cosmology will be unambiguous and therefore a
strong motivator to develop alternative theories, as well as possibly
giving us a clue as to the nature of these theories.

There have been many recent studies looking at model independent tests
of the dark energy/cold dark matter ($\Lambda$CDM) paradigm in
GR. These can be roughly split into two categories: consistency checks
of the $\Lambda$CDM assumption \cite{song09,shapiro10,acqua10} and those
that introduce new parameters to evaluate the level of deviation from
GR (see \cite{mgreview} for a complete review of the literature). 
These parametrisations and consistency checks have considered the
majority of cosmological observations: weak lensing, the CMB,
particularly through its Integrated Sachs Wolfe (ISW) effect, Baryon
Acoustic Oscillations (BAO), SN1a Supernovae luminosity distance
observations, cluster counts, and galaxy redshift and peculiar
velocity surveys. The purpose is normally to constrain both the
background expansion history and the perturbations, or growth of
structure, around the background, with some observables being
sensitive to both. The constraints are then combined. This approach
works well because the growth and expansion are determined by the same
quantities under the concordance cosmology and this is the basis for the
consistency checks. A parameter that can be used without modifying
gravity is $\gamma$, a parametrisation of the growth index
\cite{linder05,wang98}:
\begin{equation}
\frac{d\ln D(k,a)}{d\ln a}=\Omega_m(a)^{\gamma}\,.
\end{equation}
Here, $D(k,z)=\delta(k,z)/\delta(k,z=\infty)$ where $\delta$ is the
matter density contrast. For $\Lambda$CDM, $\gamma=0.55$ is a good fit
to the growth. With the modified gravity parametrisations, the
parameters often relate to the two gravitational potentials $\Psi$ and
$\Phi$ appearing in the perturbation to the Friedmann Roberston Walker
(FRW) metric. Since different observations depend on different
combinations of these potentials, combining several experiments gives
the best constraints.
 
Constraints from current data have been examined (see \cite{mgreview}
for a review) and the general conclusion appears to be that the
concordance cosmology is consistent with all of the current data.
However, the data available today does not have enough constraining
power to rule out even relatively significant modifications to gravity
and they are certainly not precise enough to distinguish {\sl between}
modified gravity theories. Work has also gone into forecasting future
constraints on a number of theories and/or parametrisations of the
modifications to GR
\cite{gbz10b,gbz09,dossett10,daniel10b,serra09,guzik10,kosow09,heavens07,acqua10}. The
consensus is that future surveys will greatly improve prospects with
the expectation that a number of theories competing with GR will be
ruled out.

Now is a good time to consider these issues as we are in an era where
the observational front is rapidly advancing in the field. Most, if
not all, of the current and future measurements can be brought to bear
on the issue of modified gravity. We have had pioneering ground and
space-based CMB experiments over the last 20 years and now await the
data from the Planck satellite \cite{planck}. Weak lensing surveys of
cosmic shear have also matured into a precise observational tool
\cite{cfhtls} and future, planned surveys promise to bring these
measurements to the fore front of the data landscape with large scale
surveys underway or in the development stage (e.g.  PANSTARRS \cite{panstarrs},
DES \cite{des}, and LSST \cite{lsst}). In addition large scale surveys
of galaxy redshifts have already been carried out (SDSS \cite{sdss},
and 2DFGRS \cite{2dfgrs}) with even larger and deeper ones targeting
BAO measurements to come. 

So far, we have only mentioned modified gravity as an alternative to
dark energy. The other alternative under consideration is whether the
assumptions of homogeneity and isotropy are justified and hence
whether using the FRW metric itself is justified. It is also important
to note that measurements of the background expansion alone cannot
distinguish between dark energy or modified gravity \cite{bert08} and it has also
been argued  that measurements of perturbations
may also suffer from this degeneracy if sufficiently complex models of
perturbed dark energy are allowed \cite{kunz06}.

In this work, we will only consider modified gravity and will look at
how well some future experiments, particularly DES and CMB
measurements and SZ cluster counts from Planck and its successors,
will combine to constrain certain modified gravity parameters that
characterise potential deviations from GR.

In this {\sl paper} we investigate how the combination of future
observations of CMB, weak lensing (WL), and cluster counts (CC) are
able to constrain the model independent parametrisation of modified
gravity theories. We restrict ourselves to the simplest form of
modified gravity with at most a linear redshift dependence in the
modification and no scale dependence. These assumptions are fairly
restrictive in terms of the physical mechanism that could underpin the
modified phenomenology but provide a simple starting point for
investigation into future constraints from different observables. In
Section~\ref{sec:gravity} we briefly review the model independent
parametrisation of modified gravity used in this work. In
Section~\ref{sec:obs} we describe the three observables used in our
forecasts. In Section~\ref{sec:forecasts} we review the Fisher matrix
formalism used in our calculations and the experimental `Stages'
considered in our forecasts. Our results are presented in
Section~\ref{sec:results} and we conclude with a discussion of the
results in Section~\ref{sec:disc}.

\section{Parametrised modified gravity}\label{sec:gravity}

There has been considerable discussion regarding the best ways to
address and parametrise deviations from GR. Of course, it is always
possible to calculate observables in the universe given a particular
modified gravity model, and this has been done for several models:
$f(R)$, DGP, Einstein Aether, TeVeS and conformal gravity have all
been tested against cosmological data. However, it is useful to
consider general kinds of deviations from GR that can then be used
as a `null-test' of the current paradigm. In other words, if there is
significant evidence that the parameters we consider differ from their
values under GR, then there is clearly a case that the current model
is wrong that is not predicated on a particular modified gravity
model.

Numerous sets of `modified gravity parameters' (MGPs) have been
suggested in the literature, see e.g. \cite{daniel10} for a partial
translation table and \cite{gbz10b} for a discussion of the differences with some of
the parametrisations.  Most of the parametrisations are
phenomenological modifications to the Einstein equations and typically
involve a parameter relating to the strength of gravity and a
parameter relating the two scalar potentials in the metric.

In this work we will use two parameters, $\eta$ and $\mu$, following
\cite{mgcamb}. The first, $\eta$, is the ratio of the two potentials,
$\eta=\Phi/\Psi$. This will be roughly equal to unity in GR unless any of
the particle species has large anisotropic stress, this is not
expected to be the case unless a significant amount of dark matter is
made up of massive neutrinos in conflict with Large Scale Structure
(LSS) data. The second, $\mu$, is a modification of
the poisson equation, and is essentially a time and space dependent
Newton's constant. Fourier expanding the spatial dependence with
wavenumbers $k$ and assuming isotropy, the modification of the
Poisson equation is as follows
\begin{equation}\label{eq:mgp}
k^2 \Psi(a,k)=-4\pi G a^2 \mu(a,k) \rho (a)\Delta(a,k) \,,
\end{equation}
where, $a$ is the FRW scale factor, $G$ is Newton's constant, $\rho$
is the background density of cold dark matter and $\Delta$ is the
gauge invariant density contrast given by 
\begin{equation}
\Delta=\delta+\frac{3aHv}{k}\,,
\end{equation}
where the cold dark matter density contrast is defined as $\delta =
\delta\rho(a,k)/\rho(a)$, $v$ is velocity of the dark matter and $H$
is the Hubble parameter.

The GR limit is recovered when the function $\mu$ is constant and
equal to unity. Note that we are using the potential $\Psi$ in the
Poisson equation since this determines the acceleration of
non-relativistic particles. These MGPs have been included as
parameters in the modified Einstein-Boltzmann code {\tt MGCAMB}
\cite{mgcamb} based on the well-known {\tt CAMB} \cite{camb} package
for integrating photon and matter perturbations. This parametrisation
can be related directly to particular models through the definition of
$\beta$ and $\lambda$ parameters \cite{bert08,mgcamb}.

Our potentials are defined as scalar perturbations of a flat, FRW
metric
\begin{eqnarray}\label{eq:pert}
g_{00}&=&-\left[1+2\Psi(\vec x , t)\right]\,,\nonumber \\
g_{ij}&=&a^2(t)\left[1-2\Phi(\vec x ,t)\right]\delta_{ij}\,,
\end{eqnarray}
where we have made the conformal Newtonian gauge choice to fix the
remaining two scalar degrees of freedom in the perturbed metric.  In
(\ref{eq:pert}), $\Psi$ is the Newtonian potential and is responsible
for the acceleration of massive particles whereas $\Phi$ is the
curvature potential, which also contributes to the acceleration of
relativistic particles.

We will consider a number of scenarios for the phenomenological
modification to the standard gravitational force. We will assume that
GR is valid up to a specified redshift. This is motivated by the
stringent conditions set on any modifications to the standard paradigm
by Big Bang Nucleosynthesis (BBN) observations of light element
abundances in the early universe and of CMB anisotropies at
recombination. Beyond the redshift where modifications take over we
will assume either that the $\mu$ and $\eta$ parameters remain
constant with a given value different from unity or that they follow a
simple time-dependence. 

This simple treatment raises two issues. Firstly, the time-dependence
of the modification introduced in (\ref{eq:mgp}) is not motivated by
any dynamical considerations and may not be self-consistent within the
framework of the underlying theory responsible for the departure from
GR. This is an unavoidable problem in the phenomenological approach
taken here \footnote{However see \cite{skordis09} for recent work in
  defining self-consistent MGPs for expansions around GR.} . Secondly
our prescription does not allow for any spatial dependence of the
modifications. This can be justified in part by requiring the simplest
modifications however all mechanisms which generate deviations from GR
will have to include some form of screening or cut--off mechanism, at
the very least, acting on scales close to solar system and below in
order to satisfy laboratory and solar system scale tests of
GR. Indeed, all mechanisms proposed so far have have some form of
explicit screening mechanism acting either universal or environment
dependent length scales. For the purpose of this exercise we shall
omit any screening mechanism, effectively assuming it acts on scales
smaller than any probed by the observations included in this
work. However if the screening scale is effective at even moderate
comoving scales of a few Mpc then we would expect some sensitivity of
our results to this assumption, particularly where we have included
cluster counts. We leave for future work the determination of the
effect of scale dependent modifications on cluster count predictions
which necessarily require the study of modified N-body simulations
\cite{inprep}. Any scale dependence will introduce structure in the
observables that, in principle, increases the degeneracies between
the MGPs and other cosmological parameters such as the spectral
tilt. However, including high redshift data such as CMB measurements
in any analysis would break the additional degeneracies and we do not
expect parameter constraints to be affected significantly if all
relevant scales are covered by the observations to sufficient
accuracy. In this work we will restrict ourselves to elucidating the
utility, or otherwise, of including cluster counts in combined
constraints on MGPs.

The background expansion history is already constrained to be close to
that of a $\Lambda$CDM model, we will therefore assume that the
modified gravity mimics the expansion history of a standard
$\Lambda$CDM setup. Having made this choice we are then left with
probes of inhomogeneity as observables that could constrain any
modifications. Throughout this work we take a fiducial
$\Lambda$CDM cosmology described by the following parameters: The
dimensionless Hubble rate in units of 100 Km s$^{-1}$ Mpc$^{-1}$, $h=0.71$, the
density of matter (baryons + dark matter) and dark energy in units of
the critical energy density, $\Omega_m=0.265$, and
$\Omega_{\Lambda}=0.735$ respectively, the optical depth to
recombination $\tau=0.088$, the amplitude of primordial, super horizon
curvature perturbations $\log (10^{10}A_s)=3.071$ at
$k=0.05\,h$ Mpc$^{-1}$ and their spectral index
$n_s=0.963$. These parameters correspond to the WMAP 7--year
  best--fit parameters \cite{wmap}. This model
yields a large scale structure normalisation of $\sigma_8=0.804$ for
the standard deviations of fluctuations on scales of $8\,h^{-1}$ Mpc.

\section{Observables}\label{sec:obs}

A number of authors have examined the use of combinations of
observables for forecasting future constraints on MGPs
\cite{gbz10b,gbz09,dossett10,daniel10b,serra09,guzik10,kosow09,heavens07,acqua10}. Combinations
have included weak lensing, CMB, galaxy redshift surveys, peculiar
velocity surveys and cluster counts.  In this work we will examine the
combination of CMB cross-correlated with weak lensing surveys and
combined with cluster counts. We have made this choice of observables
due to the inherent simplicity in their sensitivity to MGPs and the
potential to unambiguously interpret the data. The CMB on largest
scales will be sensitive to any modification of gravity through the
ISW effect as probed by photons, whilst on
the smaller scales will still provide uncorrelated constraints on the
conventional parameters of the cosmological model. Weak lensing, being
a relatively low--redshift sourced signal, will provide direct
constraints on the MGPs, also, as probed by photons. The
cross--correlation of the two observables will serve to enhance the
sensitivity to the MGPs and to reduce the degeneracies between the
MGPs and other parameters.

As with all probes of MGPs, a set of observables that depends on the
growth of non-relativistic matter perturbations is required to
constrain any difference between metric perturbations. For this
investigation we have chosen to focus on cluster counts. These offer
the prospect of an unbiased tracer of the dark matter distribution as
opposed to measurements of galaxy redshift power spectra which are
known to suffer from scale dependent biasing.

To calculate the theoretical predictions from models of modified
gravity we will employ the {\tt CAMB sources}
\footnote{http://camb.info/sources/} package which has been modified
to include modifications of gravity during the free streaming regime
after recombination ({\tt MGCAMB} \cite{mgcamb}). The {\tt CAMB
  sources} package allows the calculation of full--sky power spectra
for a number of generic `sources' which can be those of eg. weak
lensing, 21cm emission, etc. A useful feature of the package is the
capability of separating out the contribution to an individual source
into redshift bins, a feature we will be making use of for our
forecasted weak lensing surveys and cluster count predictions. All the
models used assume a standard $\Lambda$CDM background and the effect
of modified gravity is solely to change the evolution of perturbations
around the background. Additionally, We use the matter power spectrum
calculated by the code as an input for our calculation of the mass
functions, which are integrated to give the expected cluster counts
described below.

Throughout this Section and as we discuss our results we will refer to
a number of `Stages' of observations which summarise a time-line of
future surveys in each category of observables discussed. The exact
definition of the Stages considered will be discussed in
Section~\ref{sec:forecasts} after we introduce our three choices of
observables in more detail.
 
\subsection{Cluster Counts}\label{sec:clusters}

Galaxy clusters are some of the largest collapsed structures in the
universe. According to the standard $\Lambda$CDM cosmology, they
typically consist of hot gas bound in a large cold dark matter
halo. Clusters have been looked at in the context of constraining dark
energy \cite{wang98,manera06,waiz10,alam10,erlich08,battye03,basil10},
and some of the studies looking at constraining $\gamma$ or the MGPs
\cite{rapetti10,wang98,shapiro10,tang06,kosow09,schmidt09,kobay10}. They
are a useful cosmological probe as their size corresponds to scales
near the linear to non--linear transition in the underlying dark
matter power spectrum. This has several consequences: they probe the
tail of the matter perturbation spectrum and are therefore a sensitive
probe of growth. In addition, galaxy cluster counts can be predicted
accurately from linear theory, using semi-analytic formulae or
formulae calibrated from N-body simulations. These prescriptions have
been accurately calibrated using $\Lambda$CDM models but may need to
be revisited if they are to be extended to modified dark energy or
gravity models. The formulae that are calibrated by N-body simulations
work over a range of cosmologies, but their suitability to perturbed
dark energy and modified gravity cosmologies have not been
investigated fully. Some specific dark energy models have been looked
at \cite{zhao10,li10} as well as DGP
\cite{kobay10,schmidt09,schmidt09c,wyman09,wyman10} and $f(R)$ \cite{schmidt09b}. In
most cases the non-linear fitting functions and mass functions have
been found to be sufficiently accurate \cite{macci04} \footnote{See
  also \cite{rapetti10} for references regarding testing the fitting
  function from \cite{jenkins01}}.

As previously mentioned, we are not taking into account any
physical screening mechanism in this work which would naturally lead
to modifications to the mass function and any semi-analytical
prediction of cluster counts. We will look into the impact of
screening scales on the predictions in future work \cite{inprep} and
use the $\Lambda$CDM calibrated predictions for the forecasting
exercise being carried out here.

The number of clusters observable over a fraction of the sky $f_{\rm
  sky}$ and with a redshift dependent mass resolution limit $M_{\rm
  lim}(z)$ in a redshift bin  spanning the interval $z$ to
$z + \Delta z$ can be calculated by integrating the comoving number
density $dn/dM$ of objects with mass $M$
\begin{equation}
N_{\Delta z}=4\pi f_{\rm sky}\int^{z + \Delta z}_{z}dz'
\frac{dV}{dz'\,d\Omega}\int^{\infty}_{M_{\rm lim}(z')}\frac{dn}{dM}\,dM\,,
\label{clusterbins}
\end{equation} 
where $dV/dzd\Omega=r^2(z)/H(z)$ is the comoving volume at
redshift $z$ in a flat universe, with $H(z)$ the Hubble rate and
$r(z)=\int^{z}_{0}dz'/H(z')$ is the comoving distance to that
redshift. 

Much work has gone into predicting the shape of the mass function
$dn/dM$ for a given linear power spectrum starting with the
semi-analytical Press-Schechter formalism given by the mass function 
\begin{equation}
  \frac{dn}{dM}=-\sqrt{\frac{2}{\pi}}\frac{\rho}{M}\frac{d\sigma_M}{dM}\frac{\delta_c}{\sigma_M^2}\exp\left[-\frac{1}{2}\frac{\delta^2_c}{\sigma_M^2}\right]\,,
\end{equation}
where $\rho$ is the background density of dark matter today,
$\delta_c=1.686$ is a critical density contrast and $\sigma_M^2$ is the
variance of the dark matter fluctuations in a spheres of radius
$R=(3M/4\pi\rho)^{1/3}$ defined by the integral of the linear matter
power spectrum $P(k)$ over wavenumber $k$ 
\begin{equation}
\sigma^2_M=\frac{1}{2\pi^2}\int^{\infty}_{0} W^2(kR) \,P(k) \,k^2\,dk\,,
\end{equation}
with top-hat filter function
\begin{equation}
W(kR)=3\left(\frac{\sin(kR)}{(kR)^3}-\frac{\cos(kR)}{(kR)^2} \right)\,.
\end{equation}

Successive studies have yielded formulae of increasing complexity and
accuracy \cite{ps,bond91,st99a,st99b,st02}, particularly for cluster
predictions. Other approaches have used N-body calibrated empirical
formulae for the mass function. In this work we adopt the results of
\cite{jenkins01} where the mass function is expressed as
\begin{equation}\label{eq:mass}
  \frac{dn}{dM}=-0.316\frac{\rho}{M}\frac{d\sigma_M}{dM}\frac{1}{\sigma_M}\exp\left[-|0.67-\log(\sigma_M)|^{3.82}\right]\,.
\end{equation}

Since we keep an expansion history that is consistent with $\Lambda$CDM,
the effect of the modified gravity parameters will be to change the
growth history and hence the matter power spectrum. The linear matter
power spectrum is calculated at the desired redshifts by the {\tt
  MGCAMB} code, and this is then fed into the mass function
(\ref{eq:mass}) and cluster abundance
(\ref{clusterbins}). 

\begin{figure}[t]
\begin{center}
\includegraphics[width=3in]{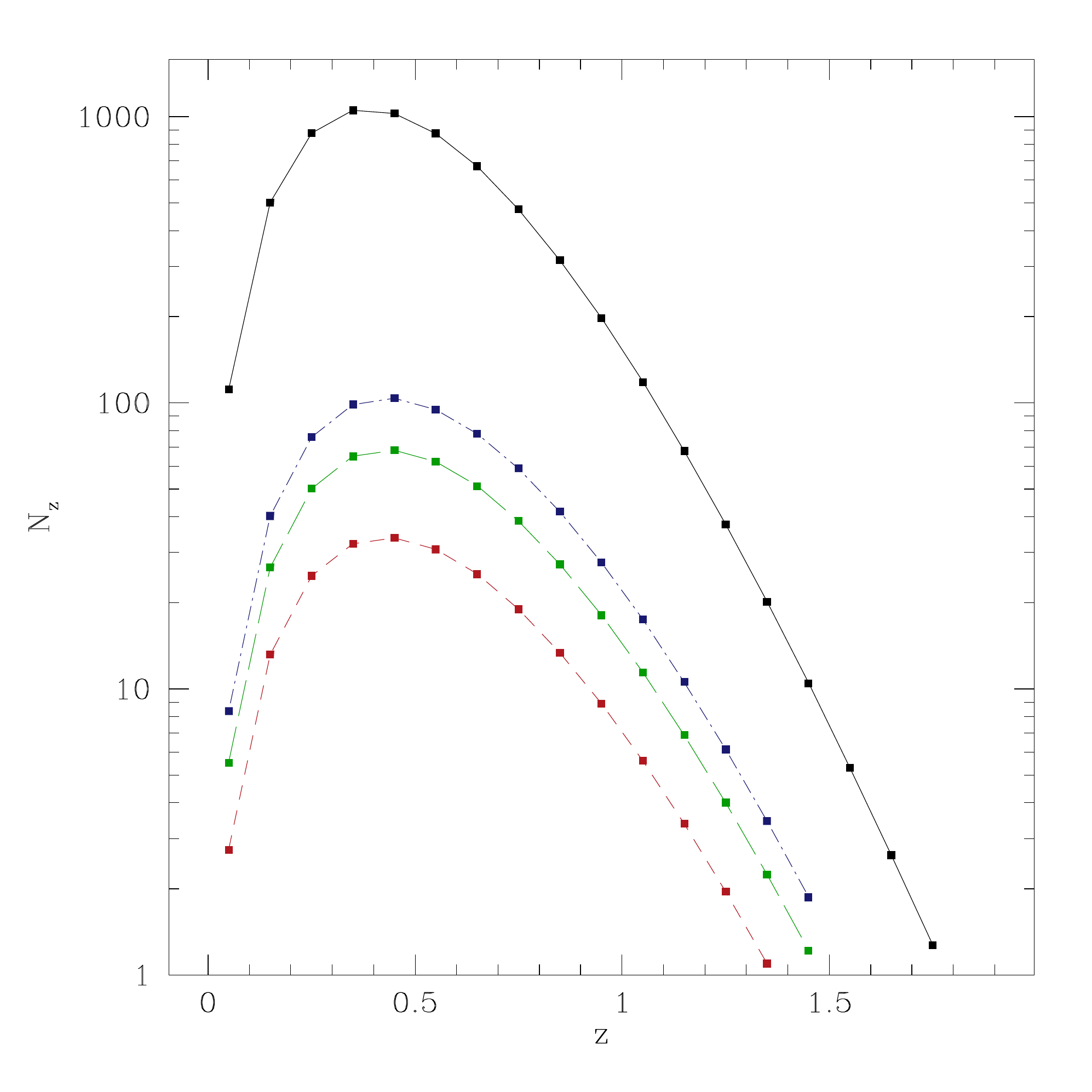}
\end{center}
\caption{Cluster counts in 20 redshift bins for the fiducial
  $\Lambda$CDM model calculated using (\ref{eq:mass}) with redshift
  independent $M_{\rm lim} = 5.0\times 10^{14} M_{\odot}$, solid
  (black). We also show the expected {\sl increase} in counts for models
  with $\mu=1.0023$ (short-dashed, red), 1.0046 (long-dashed, green),
  and 1.0069 (dash-dotted, blue) corresponding to 1, 2, and
  3$\sigma$ deviations from GR for our Stage I forecasted constraints
  (see Section~\ref{sec:results}).}
\label{fig:Ni}
\end{figure}

As can be seen in (\ref{eq:mass}), the number of clusters is
exponentially sensitive to the amount of growth that has occurred, and
should therefore provide a strong constraint on the $\mu$
parameter. This sensitivity can be seen in Figure~\ref{fig:Ni}, where
we show the expected cluster counts in redshift bins of width $\Delta
z=0.1$ for a Planck--like experiment observing 70\% of the sky. The
curve was obtained from (\ref{eq:mass}) assuming a constant limiting
mass $M_{\rm lim} = 5.0\times 10^{14} M_{\odot}$ , consistent with
what is expected of Planck \cite{planck}. We also show the increase in
cluster counts expected for the same observational setup with
$\mu=1.0023$, $1.0046$ and $1.0069$ corresponding to 1, 2, and
3$\sigma$ deviations from GR for our Stage I forecasted constraints
(see Section~\ref{sec:results}). These values represent a departure
from the GR value of $\mu$ of less than 1\%, however they lead to an
increase of between 5 and 10\% in the number of observed clusters.

Our theoretical predictions for the number of clusters in redshift
bins will be compared to predicted SZ catalogues for a number of
future observational stages. The SZ effect \cite{sz1} is a nearly
redshift independent tracer of clusters that is due to the
re-scattering of CMB photons by hot intracluster gas. The observational
limits on SZ observations are, in principle, determined simply by
resolution and sky coverage. The frequency dependence of the effect
also makes it possible to cleanly identify clusters in any
multi-frequency CMB observations spanning the null frequency of $\sim
220$GHz where the effect changes sign. An additional advantage of SZ
surveys over X-ray surveys of clusters is that the effect is less
sensitive to the internal structure of the cluster than X-ray emission
and this should make it easier to obtain unbiased estimates of the
cluster masses \cite{norman10,battye03}.

In principle, the limiting mass for an SZ survey is a redshift
dependent quantity. The source of this is the change of apparent size
of the cluster with redshift; the actual distortion to the temperature
has no redshift dependence. We calculated the effect that the full
redshift dependent limiting mass had on the constraints (see
Appendix~\ref{app:mlimz}).  Since the effect of the redshift dependent
limiting mass on the constraints turns out to be negligible, we have
used a constant, redshift independent limiting mass for all our
forecasts.

The exponential sensitivity of cluster counts to the amplitude of the
underlying density perturbations introduces some issues of
accuracy. The mass of a given cluster must necessarily be estimated
from some proxy signal such as X-ray temperature or SZ flux. Use of
cluster counts to constrain model parameters is therefore subject to
any bias introduced in the determination of the cluster mass from the
available information. Any forecasts that do not take this
uncertainty into account may potentially underestimate the errors in
model parameters. 

\begin{figure}[t]
\begin{center}
\includegraphics[width=3in]{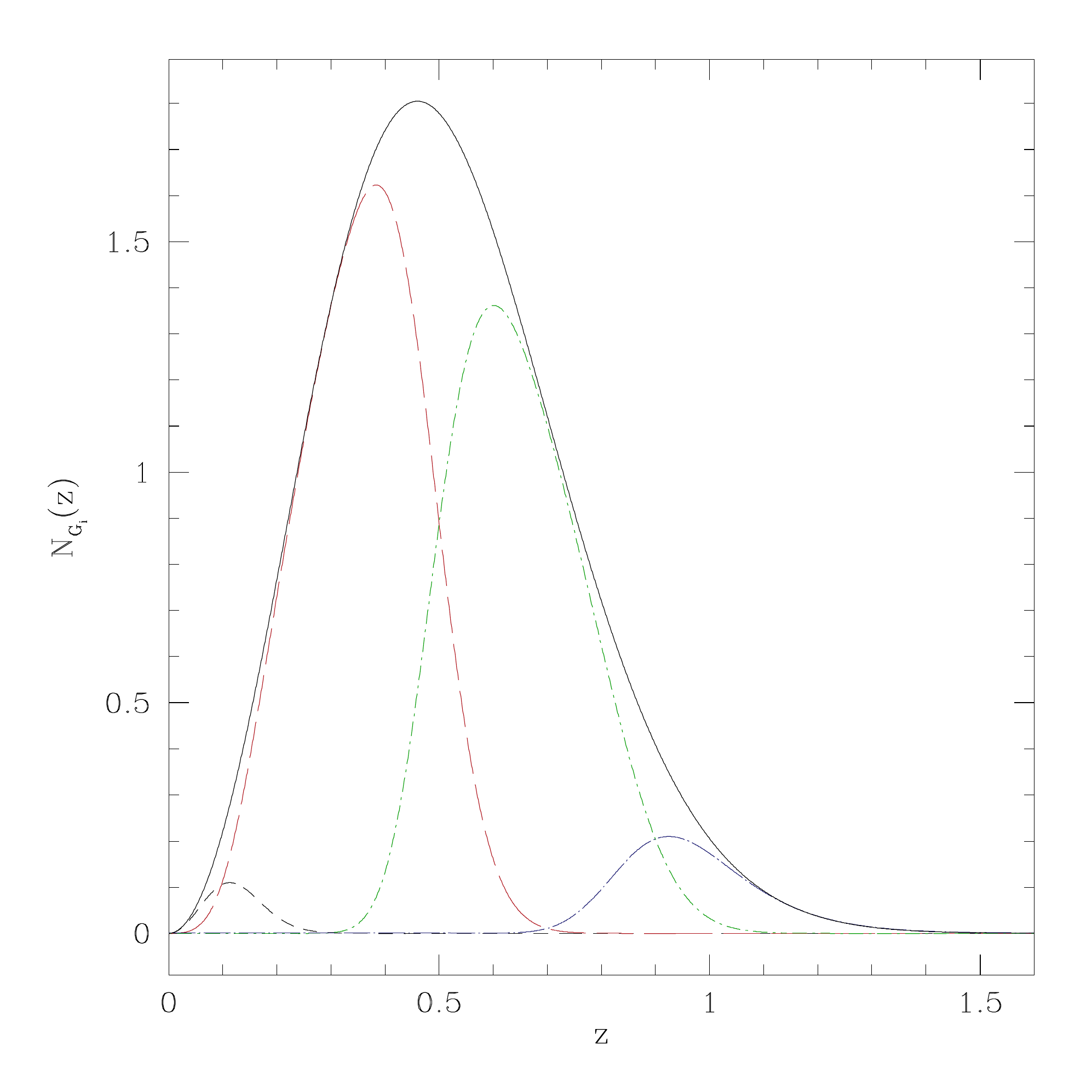}
\end{center}
\caption{Total redshift distribution of the sources (solid line) and
  redshift coverage of the four weak lensing bins used for forecasting
  the weak lensing experiments. In calculating the redshift coverage
  of the bins, we have marginalised over the uncertainty in
  determining the redshift of the sources.}
\label{fig:bins}
\end{figure}

Following \cite{limahu05}, we consider the probability of assigning a mass
$\widetilde M$ to a cluster of true mass $M$ to be given by a Gaussian
distribution in the logarithm of $M$
\begin{equation}
P(\widetilde M| M) d\widetilde M=\frac{1}{\sqrt{2}\sigma_{\cal M}}\exp\left[-\frac{1}{2}\frac{(\widetilde {\cal
  M}-{\cal M})^2}{\sigma_{\cal M}^2}\right] d\widetilde M\,,
\end{equation}
where ${\cal M}=\ln M$ and $\widetilde {\cal M}=\ln \widetilde M$. In defining the
distribution we have assumed there is no systematic bias in the
estimates of the cluster masses, only a scatter induced by the
uncertainty. The scatter is parametrised by the standard deviation of
the distribution $\sigma_{\cal M}$. 

We can now calculate the expected comoving number density given the
distribution in estimated masses with observational cutoff $\widetilde M_{\rm lim}$
\begin{equation}
n (\sigma_{\cal M}) =\int^{\infty}_{\widetilde M_{\rm lim}}\frac{d\widetilde M}{\widetilde M}\int^{\infty}_{0} dM\frac{dn}{dM}\,P(\widetilde M\vert M)\,.
\end{equation}
Substituting in the expression for the probability and carrying out
the integral yields
\begin{equation}\label{eq:selfcal}
n(\sigma_{\cal M}) =\frac{1}{2}\int^{\infty}_{0} dM\frac{dn}{dM}\,{\rm
  erfc}\left[\frac{1}{\sqrt{2}}\frac{(\widetilde {\cal
  M}_{\rm lim}-{\cal M})}{\sigma_{\cal M}}\right]\,.
\end{equation}
Integrating the density over redshifts gives a modified number count
as a function of the uncertainty $\sigma_{\cal M}$
\begin{equation}
N_{\Delta z}(\sigma_{\cal M})=4\pi f_{\rm sky}\int^{z + \Delta z}_{z}dz'\,n(\sigma_{\cal M})\,\frac{dV}{dz'\,d\Omega}\,,
\end{equation}
This can be used in our forecasts to account for the uncertainty. We
will adopt a reference value of $\sigma_{\cal M}=0.25$
\cite{limahu05}. We will include the  $\sigma_{\cal M}$ as an extra
parameter in the Fisher matrices for the cluster counts and marginalise
it out by eliminating its contribution to the inverse Fisher matrix.

\subsection{CMB}
With the release of Planck satellite results only a few years away we
are entering an era where observations of the CMB total intensity
spectrum will have reached the sample variance limit throughout scales
where primary effects dominate the signal. As such, in all of our forecasts, we
will be assuming a Planck--like observation of the CMB angular power
spectrum out to multipoles $\ell=2000$.

The sensitivity to MGPs in the CMB spectrum is restricted to the
largest scales. This is due the constraint that gravity is not
modified at early times and through recombination. This means
anisotropies generated through the Sachs Wolfe effect on super-horizon
scales and acoustic effects on sub-horizon scales at recombination will
not be affected by our late time modifications. The only signal will
arise on the largest scales due to the ISW effect which is sourced as
the Universe transitions into a dark energy dominated model and the
potential starts to decay. The effect can be described by the integral
of the time-derivative of the sum of metric potentials along the line
of sight as photons free stream after recombination
\begin{equation}
  \left(\frac{\delta T}{T}\right)_\ell = -\int_0^{\tau_0}
  e^{-\theta(\tau)}\frac{\partial}{\partial \tau}\left[\Phi(1+\eta)\right] \,j_\ell[k(\tau-\tau_0)]\,d\tau\,,  
\end{equation}
where $(\delta T/T)_\ell$ is the multipole expanded, fourier transform
of the CMB temperature fluctuation at conformal time today, $\tau_0$,
$\theta$ is the optical depth, and the spherical Bessel functions
$j_\ell$ describe the projection of plane--wave modes on the celestial
sphere. The ISW contributes to the power on the largest
scales as it is only sourced at late times (we will disregard any
effect from the early ISW effect due to potential evolution close to the
radiation to matter transition). These modes are fundamentally
ill--sampled due to the small number of $a_{\ell m}$ coefficients on
these scales and also suffer degeneracy with other effects such as
the Sachs Wolfe signal from last scattering and the effect of
reionisation. One way to gain further constraining power from such a
sample variance limited signal is to cross-correlate the CMB with other
large scale observables, indeed, low-significance detections of a dark
energy component have been reported via cross--correlation with a
number of tracer of large scale structure
\cite{crit96,cooray02,boughn04,afshord04,cora05,gian06,piet06,gian08,cabr06}.
We will also take advantage of this by cross-correlating with template
weak lensing surveys.

As our template CMB observable we take an angular power spectrum
$C_\ell$ from the best--fit $\Lambda$CDM model and add sample and
noise variance according to a Planck--like survey covering an area
corresponding to 70\% of the sky. The error at each multipole can be
calculated from the observational parameters via
\begin{equation}\label{eq:cmberr}
\delta C_{\ell}=\sqrt{\frac{2}{f_{\rm sky}(2\ell+1)}}\left(C_{\ell}+N_{\ell}\right)\,,
\end{equation}
where $N_{\ell}$ is a function of resolution 
($\theta$), the number of detectors ($N_{\rm det}$) and Noise Equivalent
Temperature ($T_{\rm NET}$) in a given channel, as well as the overall
integration time ($t_{\rm tot}$)
\begin{equation}
N_{\ell}=\left[\frac{1}{\left(\sigma
      \theta\right)^2}\exp\left(-\frac{\ell(\ell+1)\theta}{8\ln 2}\right)\right]^{-1}\,,
\end{equation}
with $\sigma$ evaluated as
\begin{equation}
\sigma=\frac{4\pi
  f_{\rm{sky}}}{\theta}\frac{T_{\rm NET}}{\sqrt{N_{\rm det}t_{\rm tot}}}\,.
\end{equation}
For our Planck--like survey, we will use the values $\theta=7.1$
arcminutes, $N_{\rm det}=32$, $T_{\rm
  NET}=62.0$ $\mu$K s$^{1/2}$, $f_{\rm sky}=0.7$ and $t_{\rm tot}=14$ months.

\subsection{Weak lensing}
\label{section:wl}
The third observable we will use is the convergence power spectrum
from weak lensing surveys. Weak lensing is a relatively new
cosmological tool and is a measure of the small distortions of
background galaxies due to gravitational lensing by large scale
structure \cite{kaiser92}. Distortions of
individual background galaxies are virtually impossible to measure due
to the intrinsic ellipticity of galaxies. However, statistical results
averaging over large numbers of galaxies are now routinely reported
\cite{refregier03}. The convergence $\kappa$ is a measure of the Laplacian of
the potentials responsible for the lensing along the line of sight and
can be calculated as an integral over the comoving radial distance $r$
as
\begin{equation}
  \kappa = \frac{k^2}{2} \int^{r_\infty}_0  \Phi(1+\eta)\, g(r)dr\,,
\end{equation}
where we have expressed quantities in the Fourier domain and $g(r)$ is
a filter function determined by the redshift distribution of background
galaxies being lensed $w[r(z)]$ 
\begin{equation}
  g(r) = r \, \int^{r_\infty}_0 \left[1-\frac{r'}{r}\right]w(r')\,dr'\,.
\end{equation}
The convergence power spectra can be calculated using {\tt MGCAMB} for
the modified gravity case and can be split into contributions from
separate redshift bins assuming the observations are able to obtain
sufficiently accurate photometric redshifts of the background
sources. In all cases we will include multipoles $\ell \le 2000$ to
avoid complications that arise due to lensing from non-linear scales.

\begin{table}[t]
\caption{Parameters used for the three Stages of future observations
  used in our forecasts.}  
\centering      
\vskip 0.5cm
\begin{tabular}{|l |c| c| c| c|}
  \hline                   
  & f$^{\kappa}_{\rm{sky}}$ &   N$_{g}$   &   f$^{\rm{SZ}}_{\rm{sky}}$   &   M$_{\rm{lim}}$ (M$_{\odot}$) \\ [0.5ex]   
  \hline  
  Stage I & 0.121 &    $2.14\times10^{8}$ &    0.7 &  $5.0\times10^{14}$ \\   
  Stage II & 0.485 &    $3.6\times10^{8}$ &    0.7 &  $2.5\times10^{14}$\\
  Stage III & 0.485 &    $2.88\times10^{9}$ &    0.7 &  $1.0\times10^{14}$ \\[1ex] 
  \hline  
\end{tabular}
\label{table:3layers}
\end{table}

For our initial weak lensing survey, we consider a DES-like
survey. DES is a ground based survey at the Cerro-Tololo
Inter-American Observatory in Chile that is scheduled to begin
observations in 2011. It will survey 5000 sq deg over 5 years and aims
to constrain dark energy with 4 probes: supernovae, BAO, galaxy
clusters and weak lensing, the latter being the probe we are
interested in here. We consider 4 redshift bins between $z=0$ and
$z=2$, following the prescription in \cite{gbz10b}. 

One of the biggest sources of error in these surveys will be errors in
the photometric determination of redshifts of the background
galaxies. We model the overall redshift distribution of the sources as
\begin{equation}
  w(z) = N_g \frac{4}{\sqrt{\pi}} \frac{z^2}{z_\star^3} \exp\left[ -\frac{z^2}{z_\star^2}\right]\,,
\end{equation}
such that $\int w(z) \, dz = N_g$, the total number of background
galaxies and $z_\star$ defines the median redshift of the
distribution. We take a reference value of $z_\star=0.46$ for our
template weak lensing surveys.

We can take into account the uncertainty in photometric redshifts when
breaking down the signal into contributions from different redshift
bins. We define four, overlapping distributions $w_i(z)$ with
$i=$1, 2, 3, and 4 with the constraint
\begin{equation}
  w(z) = \sum_{i=1}^4 w_i(z)\,,
\end{equation}
with 
\begin{equation}
  w_i(z) = \frac{1}{2}w(z)\left[{\rm
      erfc}\left(\frac{z_{i-1}-z}{\sqrt{2}\,\sigma(z)} \right)-{\rm erfc} \left(\frac{z_{i}-z}{\sqrt{2}\,\sigma(z)} \right) \right]\,.
\end{equation}
The bins are centred at redshifts $z_i$ corresponding to 0.1, 0.5,
0.9, and 1.3, as shown in Figure~\ref{fig:bins}, with the photometric redshift error given by 
$\sigma=0.05(1+z)$.

\begin{figure}[t]
\begin{center}
\includegraphics[width=3in]{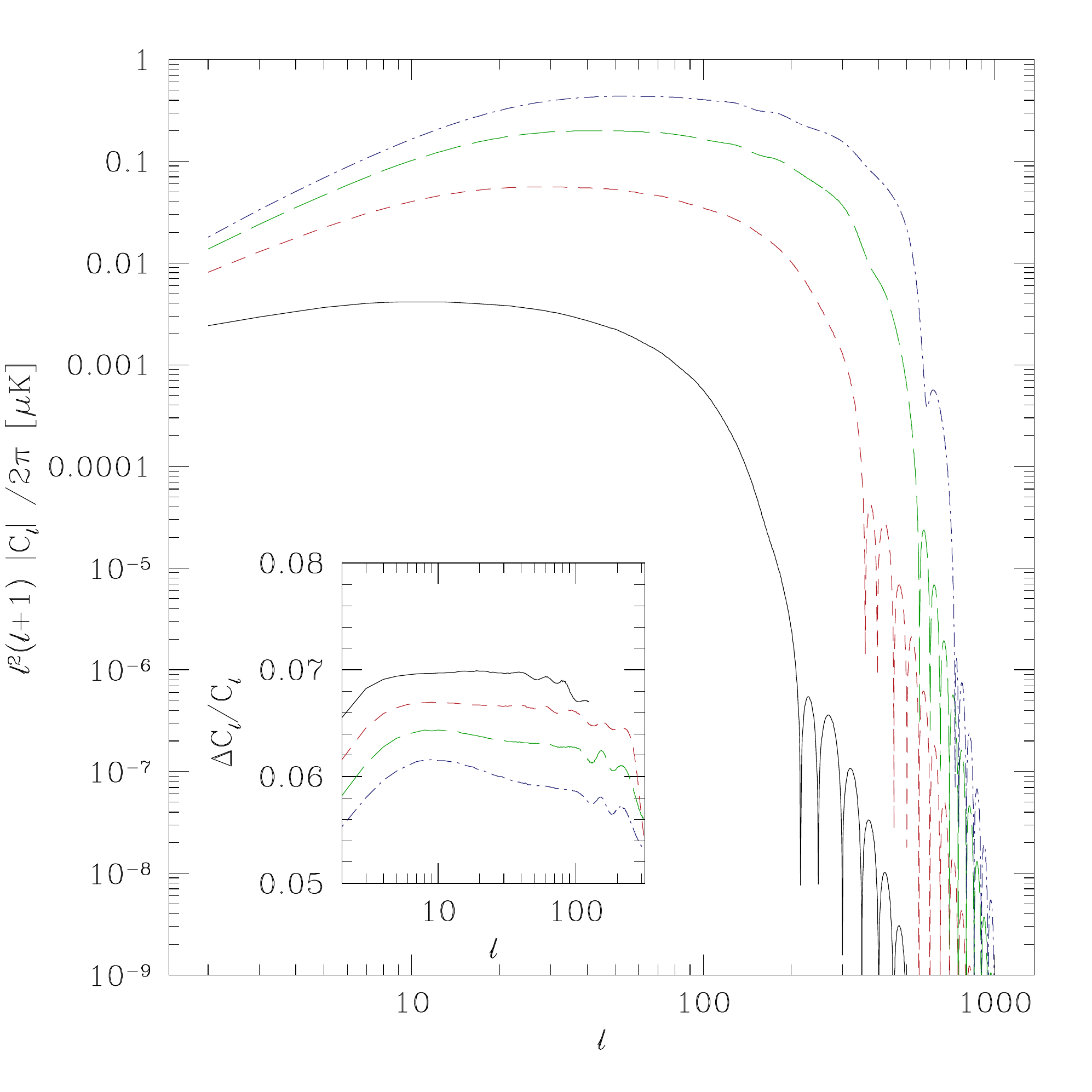}
\end{center}
\caption{The cross--correlation signal between CMB temperature and weak
  lensing convergence from Stage I type survey combination (see
  Table~\ref{table:3layers}) as a function of angular multipole
  $\ell$. The curves shown are for the fiducial $\Lambda$CDM
  model. The cross--correlation is shown in four redshift bins with
  amplitude increasing with redshift. The inset shows the fractional
  change in the cross--correlation signal with MGP values $\eta=1.03$
  and $\mu=1.008$ corresponding to a 1$\sigma$ shift given our final
  constraints (see Section~\ref{sec:results}).}
\label{fig:cl}
\end{figure}

We can model the statistical error in the angular power spectrum of
the convergence from each redshift bin as a sum of sample and noise
contributions
\begin{equation}\label{eq:wlerr}
  \delta C_\ell^{\kappa_i} = \sqrt{\frac{2}{f_{\rm
        sky}(2\ell+1)}}\left( C_\ell^{\kappa_i} + \frac{\langle
      \gamma^2 \rangle}{N^i_g}\right)\,,
\end{equation}
where $\langle \gamma^2 \rangle$ is the variance of the intrinsic
ellipticity of a typical galaxy in the survey and $N^i_g=\int w_i(z)\,
dz$ is the number of galaxies in each redshift bin. See table \ref{table:3layers} for the
parameters used to model the three Stages of weak lensing experiments. In all cases
$\langle \gamma^2\rangle=0.16$ and the photometric redshift error is given by
$\sigma=0.05(1+z)$.

With the advent of large scale weak lensing surveys the possibility of
cross-correlating CMB and convergence maps will become a reality. This
will make use of all the available information in the data since the
signal in two such maps will be correlated. The cross--correlation will
be most useful in this case since dependence on MGPs is expected to be
strongest on the largest scales. In Figure~\ref{fig:cl} we show the
expected cross--correlation signal between CMB and the four weak lensing survey
redshift bins used in our analysis. The inset shows the fractional
change in the cross--correlation induced by a change in MGPs from their
fiducial GR values to $\eta=1.03$ and $\mu=1.008$. These correspond to 1$\sigma$
deviations from the fiducial values given our forecasted constraint
obtained from CMB and weak lensing data alone (see Section~\ref{sec:results}).

\begin{figure}[t]
\begin{center}
\includegraphics[width=3.2in]{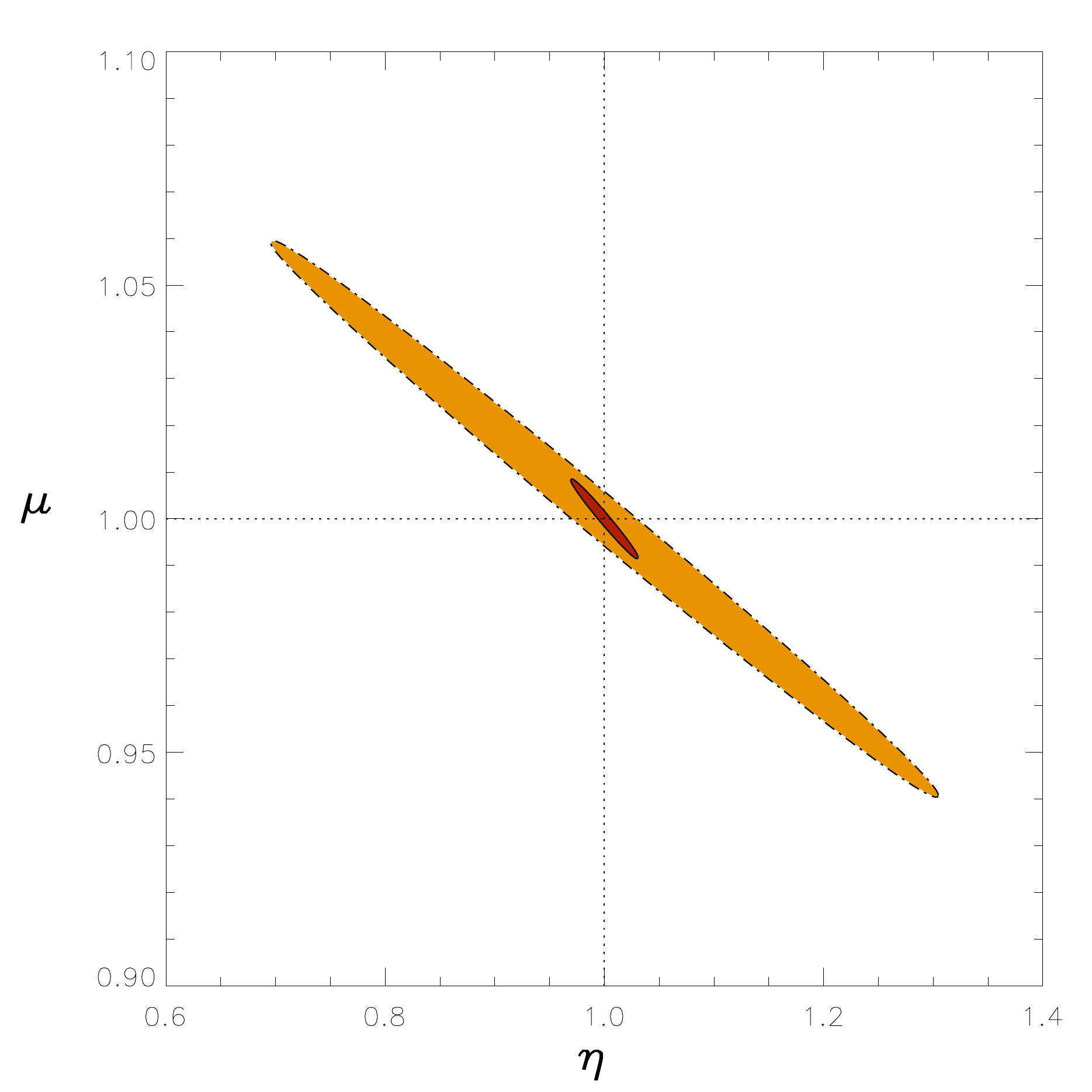}
\end{center}
\caption{Fisher constraints on $\eta$ and $\mu$ from the combination
  of Stage I CMB and weak lensing observations with and without
  cross--correlation (orange/dot-dashed ellipse and red/solid ellipse
  respectively).}
\label{fig:pre-cross}
\end{figure}
\begin{figure}[t]
\begin{center}
\includegraphics[width=3.2in]{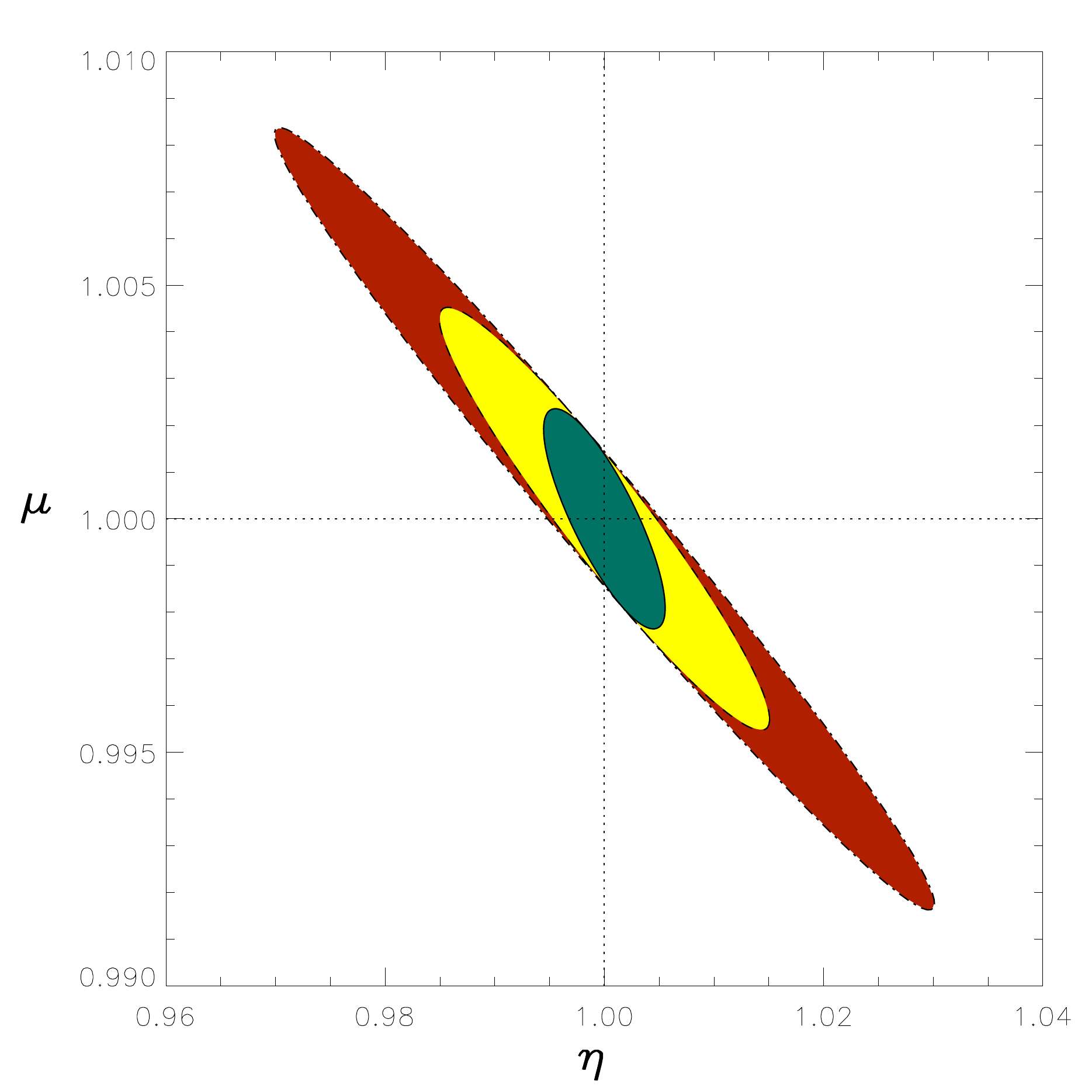}
\end{center}
\caption{Fisher constraints on $\eta$ and $\mu$ from cross-correlated
  CMB and weak lensing measurements (Stage I) are shown by the red
  area (dot-dashed contour). The improvement obtained by adding Stage I
  cluster counts is seen in the green area (solid contour). The outer
ellipse corresponds to the inner ellipse of
Fig.~\ref{fig:pre-cross}. Adding an uncertainty in the mass assignment
for the clusters of $\sigma_{\cal M}=0.25$ decreases the impact of
adding cluster data as shown by the yellow (dashed) ellipse.}
\label{fig:tjenk1}
\end{figure}

\section{Forecasts}\label{sec:forecasts}

In this Section we carry out forecasts for a number of future
observational `Stages'. Since the Planck satellite will provide a
sample variance limited map of CMB total intensity anisotropies
covering angular scales where the signal of interest lies we will use
our Planck--like setup as the CMB contribution throughout. For weak
lensing and cluster counts we will assume three distinct observational
stages corresponding to short, medium, and long--term development of
survey sizes and accuracies.

\begin{itemize}
  \item {\bf Stage I}: Corresponds to a Planck-like cluster survey
and a DES (Dark Energy Survey)-like weak lensing survey. DES is being
carried out on the Cerro Tololo Inter-American Observatory in the
Chilean Andes and should start taking data in late 2011. DES is modelled
with a total of 214 million galaxies over 12\% of the sky.
\item {\bf Stage II}: This stage includes a weak lensing survey based
  on PANSTARRS and as such is modelled with a total of 360 million
  galaxies over 48.5\% of the sky. The SZ survey is modelled by
  keeping the same $f_{\rm sky}$ as with Planck, but lowering the
  limiting mass to $2.5\times10^{14}M_{\odot}$.  This corresponds to
  lowering the smallest change in flux that the SZ survey can detect.
  \item {\bf Stage III}: This includes a weak lensing survey based on the
LSST, due to begin taking data in 2020. This survey is modelled with the
same fraction of the sky as PANSTARRS (48.5\%), but with a total number
of galaxies of 2880 million. The third stage SZ survey assumes a
limiting mass of $1.0\times10^{14}M_{\odot}$.
\end{itemize} 

The sampling characteristics for the three Stages are summarised in
Table~\ref{table:3layers}. We have kept the same intrinsic
ellipticity, redshift bins and photometric errors assumptions for all
Stages as it is unclear how photometric redshift resolution will
evolve as surveys increase in size and complexity.

\begin{table*}[t]
\centering
\begin{tabular}{|c| c| c |c|c|c|c|}
\hline
&\multicolumn{2}{|c|}{Stage I}&\multicolumn{2}{|c|}{Stage 
II}&\multicolumn{2}{|c|}{Stage III} \\
\hline 
Parameter & CMB $\otimes$ WL & CMB $\otimes$ WL + CC& CMB $\otimes$ WL & 
CMB $\otimes$ WL + CC& CMB $\otimes$ WL & CMB $\otimes$ WL + CC\\ 
[0.5ex]
\hline
$\Delta\eta$ & $3.0\times 10^{-2}$ & $5.5 \times 10^{-3}(1.5\times 
10^{-2})$& $1.8 \times 10^{-2}$  & $4.4 \times 10^{-3} (8.8\times 
10^{-3})$ & $1.5 \times 10^{-2}$  & $2.4 \times 10^{-3} (5.5\times 
10^{-3})$ \\
$\Delta\mu$  & $8.4\times 10^{-3}$ & $2.3\times 10^{-3}(4.5\times 
10^{-3})$ & $5.5\times 10^{-3}$ & $2.1 \times 10^{-3}(3.1 \times 
10^{-3}) $& $4.8\times 10^{-3}$ & $1.5 \times 10^{-3}(2.1 \times 
10^{-3}) $ \\[1ex]
\hline
\end{tabular}
\caption{Fisher errors on $\eta$ and $\mu$ for Stage I, II, and III 
observations for different
   combinations of observables: cross--correlation of CMB and weak
   lensing (CMB $\otimes$ WL) and addition cluster counts (CC). The
   values in brackets show how the errors are affected by the
   marginalisation over a mass assignment error.} \label{table:jenk1}
\end{table*}

Our forecasts are based on Fisher matrix estimates of errors in a
subset of parameters that comprises the MGPs, $\eta$ and $\mu$, and the two
parameters from the standard model that are expected to be most
correlated with them, the total matter density $\Omega_m$ and
the primordial amplitude of scalar curvature perturbations $A$. In
varying only these four parameters we are assuming that the remaining
four standard $\Lambda$CDM parameters are well constrained by signals
orthogonal to those being used in our analysis. A combination of high resolution CMB spectra, including
polarisation $E$-modes, and a standard prior on the value of the
Hubble rate will ensure most of the remaining parameters are fixed to
within a few percent of their nominal values, which should have a minimal
impact on the four parameters being considered in this work.

An alternative to our Fisher matrix based method would be to Monte
Carlo Markov Chain (MCMC) sample the joint posterior distribution of
our parameter set by evaluating the likelihood explicitly. However
this would involve the definition of a likelihood as a function of
realisations of the observables (and their cross--correlation) and the
added complexity is not warranted for this kind of exercise at this stage.

It should be noted that use of a Fisher matrix to estimate parameter
errors assumes that the observables are distributed as Gaussian
variates. This is not true in all cases considered here since we are
considering power spectra and number counts but as long as the true
answer lies close to our fiducial values for the four parameters the
errors should give a good indication of the constraints.

Given a set of $n$, uncorrelated  measurements $C_a$, with $a=1,...,n$ and
measurement errors $\delta C_a$, the Fisher matrix for a set of $m$
parameters $\lambda_\alpha$ with $\alpha=1,...,m$, can be evaluated as
\begin{equation}
F_{\alpha \beta}=\displaystyle\sum^{n}_{a=1}\frac{1}{\delta C_a}\frac{\partial C_{a}}{\partial \lambda_{\alpha}}\frac{\partial C_{a}}{\partial \lambda_{\beta}}\frac{1}{\delta C_a}\,,
\end{equation}
where $\beta=1,...,m$.

The Fisher matrix represents the ensemble average of the negative
curvature in the log likelihood of the model parameters and its
inverse, in this limit, is therefore the covariance matrix in those
parameters. The Fisher matrix is simple to evaluate since it involves
only the first derivatives of the signal with respect to the model
parameters. These can be evaluated either analytically or
numerically. Here we use a central difference scheme to numerically
approximate the derivatives to second order in the step-size. The central difference is
sampled by evaluating the models with given step-sizes either side of
the fiducial model in all parameter directions.

For the cluster counts case the measurement consists of counts in each
of twenty redshift bins $N_i$, as shown in Figure~\ref{fig:Ni} for the
fiducial model, and the Fisher matrix is calculated as
\begin{equation}\label{eq:fish1}
  F_{\alpha \beta}=\displaystyle\sum^{20}_{i=1}\frac{1}{\sigma_{N_i}^2}\frac{\partial N_i}{\partial \lambda_{\alpha}}\frac{\partial N_i}{\partial \lambda_{\beta}}\,,
\end{equation}
where we have assumed a shot noise model for the error in the counts.

Fisher matrices from independent data can be added and then inverted
to obtain error estimates for the combination of data. Since we will
rely heavily on the cross--correlation of CMB and weak lensing
measurements to extract the relevant signal we calculate the combined
Fisher matrix for this cross--correlation to add to (\ref{eq:fish1}).

The Fisher matrix formalism can be easily extended to the correlated
measurement case. We treat the combination of CMB and weak lensing
measurements at a given multipole $\ell$ as a matrix, ${\mathbf
  C}_\ell$, of angular, cross--correlation power spectra with dimension
$n\times n$. In our case the index $n$ spans both CMB and convergence
angular power spectrum measurements (over the four redshift bins)
i.e. $T,\kappa_1,\kappa_2,\kappa_3,\kappa_4$ with the symmetric form
\begin{equation}
  {\mathbf C}_\ell \equiv \left(\begin{array}{cccc}
    C_\ell^{TT}&C_\ell^{T\kappa_1}&C_\ell^{T\kappa_2}& ...\\
    .&C_\ell^{\kappa_1\kappa_1}&C_\ell^{\kappa_1\kappa_2}& ...\\
    .&.&C_\ell^{\kappa_2\kappa_2}& ...\\
    .&.&.&...
  \end{array}\right)\,.
\end{equation}
Each measurement matrix will have a corresponding covariance matrix
with non-zero off-diagonal contributions due to the correlations in
the signal part of the measurements whilst the diagonals will have
contributions from both signal and noise
\begin{equation}
  {\mathbf \Sigma}_\ell \equiv \left(\begin{array}{cccc}
    (\delta C_\ell^{TT})^2&(\delta C_\ell^{T\kappa_1})^2&(\delta C_\ell^{T\kappa_2})^2& ...\\
    .&(\delta C_\ell^{\kappa_1\kappa_1})^2&(\delta C_\ell^{\kappa_1\kappa_2})^2& ...\\
    .&.&(\delta C_\ell^{\kappa_2\kappa_2})^2& ...\\
    .&.&.&...
  \end{array}\right)\,,
\end{equation}
where the diagonal elements correspond to the square of standard deviations
(\ref{eq:cmberr}) and (\ref{eq:wlerr}) and the off-diagonal terms can
be evaluated using
\begin{equation}
  (\delta C_\ell^{XY})^2 = \frac{1}{(2\ell+1)}\left[\frac{C_\ell^{XX}C_\ell^{YY}}{\sqrt{f_{\rm sky}^X\,f_{\rm sky}^Y}} +  \frac{(C_\ell^{XY})^2}{{\rm min}(f_{\rm sky}^X,f_{\rm
        sky}^Y)\, }\right]\,.
\end{equation}
Here $C_\ell^{XY}$ is the model cross--correlation power spectrum for
the two observables which is also computed by {\tt CAMB sources}.

The Fisher matrix for this generalised case can be evaluated as
\begin{equation}
  F_{\alpha \beta} = \sum_\ell {\rm Tr} \left[ \frac{\partial
      {\mathbf C}_\ell}{\partial \alpha}\cdot {\mathbf \Sigma}_\ell^{-1}\cdot \frac{\partial
      {\mathbf C}_\ell}{\partial \beta}\right]\,.
\end{equation}
A further contribution to the Fisher matrix is given by the sample
variance depending on the parameters
\begin{equation}
 F_{\alpha \beta} =\frac{1}{2} \sum_\ell {\rm Tr} \left[ \frac{\partial
      {\mathbf \Sigma}_\ell}{\partial \alpha}\cdot {\mathbf \Sigma}_\ell^{-1}\cdot \frac{\partial
      {\mathbf \Sigma}_\ell}{\partial \beta}\cdot {\mathbf \Sigma}_\ell^{-1}\right]\,.
\end{equation}
This term is sub-dominant to the first, however we have included it in
the analysis for the sake of completeness.

\section{Results}\label{sec:results}

\begin{figure}[t]
\begin{center}
\includegraphics[width=3.2in]{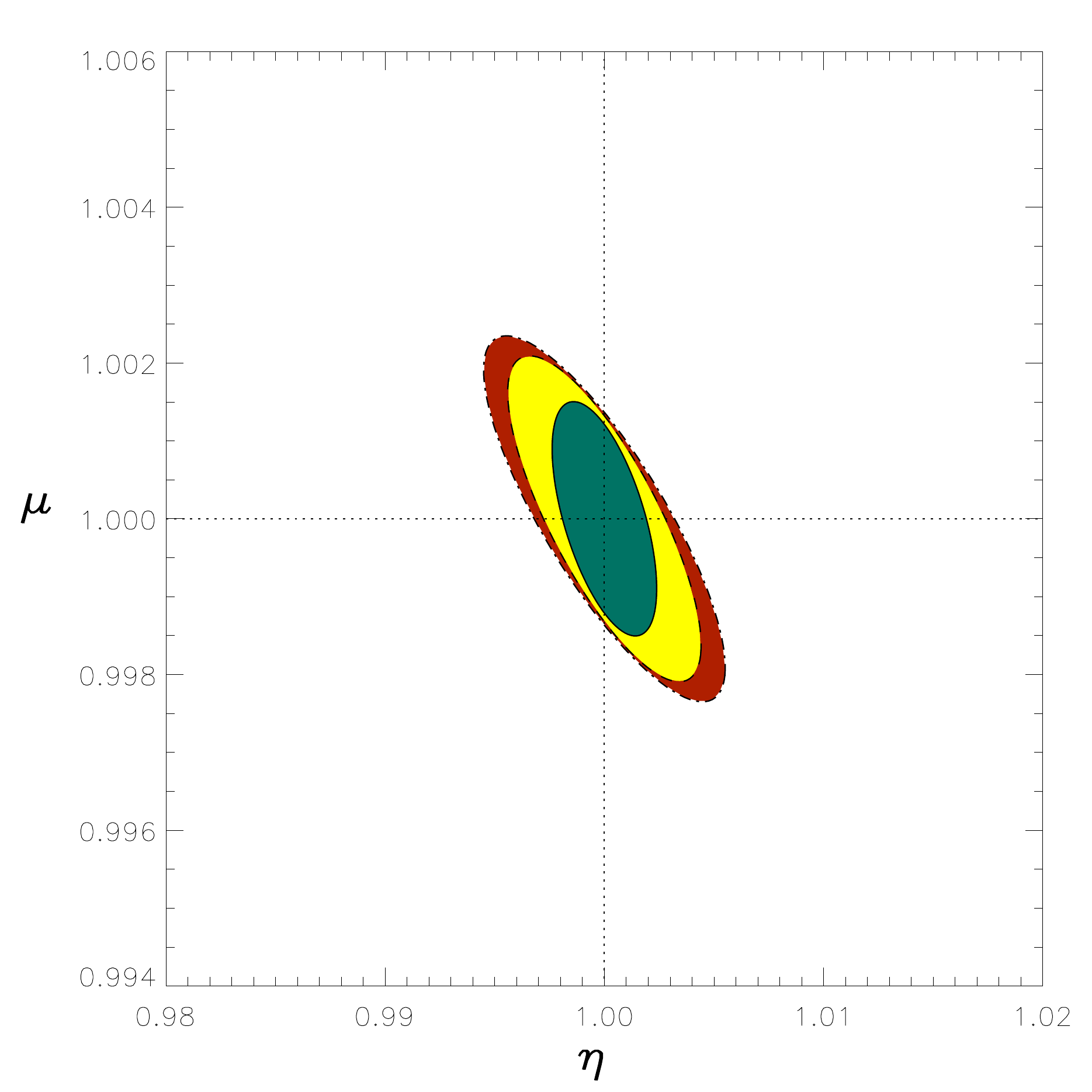}
\end{center}
\caption{Fisher constraints on $\eta$ and $\mu$ from the combination
  of CMB, weak lensing, and cluster counts for the three observational
  stages. Stage I, red (dot-dashed); Stage II, yellow (dashed); Stage
  III, green (solid).}
\label{fig:threestage}
\end{figure}

\begin{figure}[t]
\begin{center}
\includegraphics[width=3.2in]{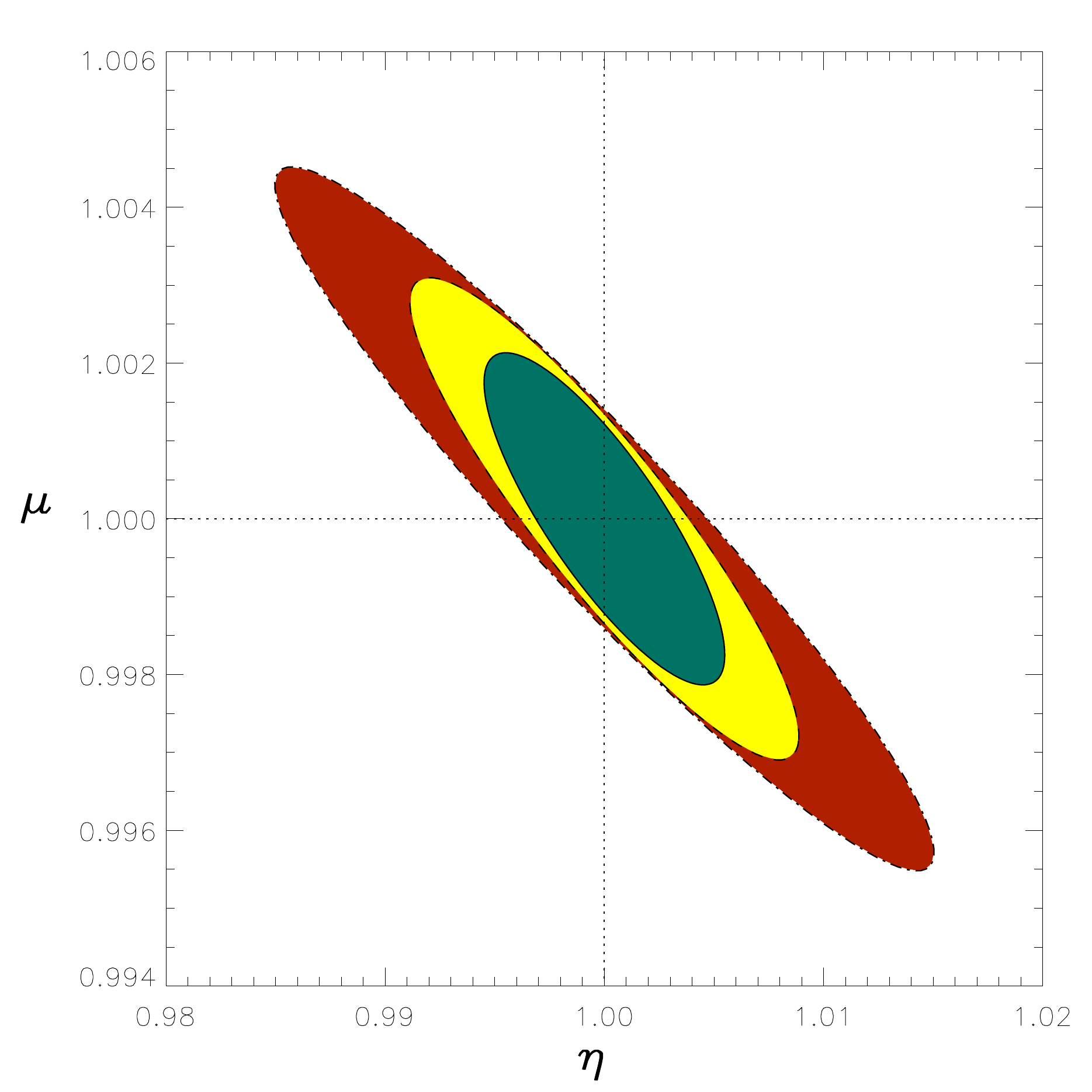}
\end{center}
\caption{Same as in Fig.~\ref{fig:threestage} but with a mass assignment
uncertainty of $\sigma_{\cal M}=0.25$ included in the constraints.}
\label{fig:jenk3}
\end{figure}

\begin{figure}[t]
\begin{center}
\includegraphics[width=3.2in]{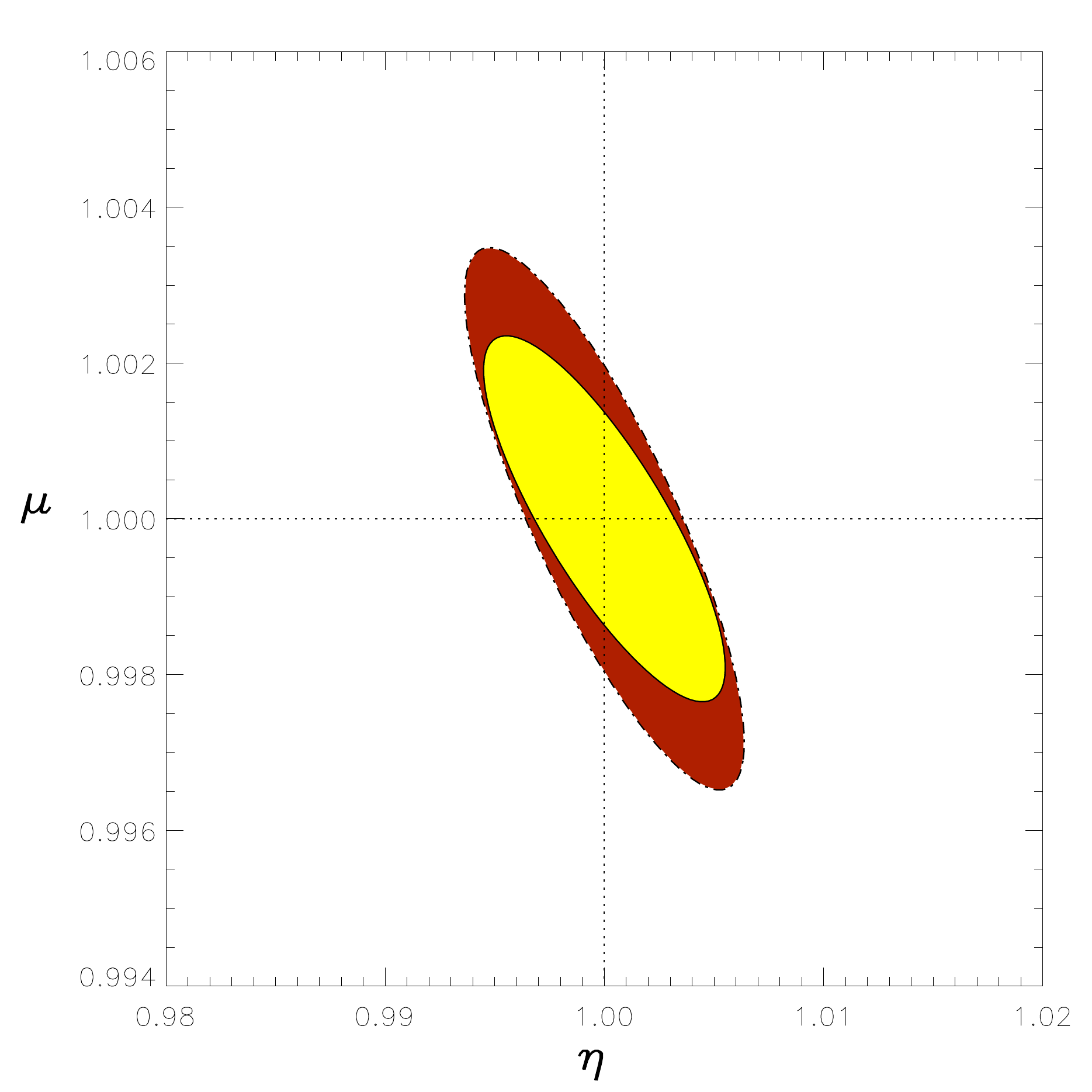}
\end{center}
\caption{Fisher constraints on $\eta$ and $\mu$ from the combined CMB,
  weak lensing, and cluster counts for the time--independent
  (solid/yellow) and time--dependent (dot-dashed/red) modified gravity
  parameters. In both cases $z_{\rm mg}=30$ but for the latter a
  linear redshift-dependence is assumed (\ref{eq:timedep}).}
\label{fig:td_jenk1}
\end{figure}

We initially considered the case where the MGPs switch from their
fiducial, GR value to a new, time independent value, at a fixed
redshift, $z_{\rm mg}$. There is a wide range in redshifts, spanning
from the time of recombination to the end of the so called
`dark-ages', where modifications to GR could come into effect. Here we
chose $z_{\rm mg}=30$ for the time-independent case which ensures that
most non-linear structures observed formed in the presence of the
modifications. We also consider the case where the MGPs have a simple,
linear redshift dependence between their GR values at $z_{\rm mg}$ and
their late time values $\eta_0$ and $\mu_0$ at $z=0$. Specifically we
show how our results change for a choice $z_{\rm mg}=3$ and $z_{\rm
  mg}=1$.

 The CMB is most effective in constraining the standard parameters of
 the concordance model of cosmology. The only effect of the MGPs is on
 the ISW effect, and this has several shortcomings. The increase of
 power due to the MGPs is not large, and this is compounded by the
 higher cosmic variance on these scales. The uncertainties in weak
 lensing mostly come down to a degeneracy between $\eta$ and
 $A$. This, is why, when combining CMB and Weak lensing observations,
 the constraints on the modified gravity parameters improve
 significantly. The extra information obtained from cross correlating
 the two observations emphasises this complementarity of the two
 sets of data. Figure \ref{fig:pre-cross} shows how the constraints on
 $\eta$ and $\mu$ improve when the cross--correlation between CMB and
 weak lensing is included in the Fisher estimates.  The contours shown
 the 1-$\sigma$ constraints obtained from the inverse of the Fisher
 matrix for all four parameters. We ensure our final errors account
 for marginalisation over our nuisance parameters $\Omega_m$
 and $\ln A$ by only ignoring the matrix elements  corresponding to
 these parameters  after inverting the full Fisher matrix.

 Even after taking into account the cross--correlation between CMB and
 weak lensing a significant degeneracy in $\eta$ remains due to the
 correlation between the two gravitational potentials in both signals.
 Adding cluster counts to the mix improves the situation since their
 signal is highly sensitive to just $\Psi$ and therefore to the
 parameter $\mu$, which affects the growth rate of structure. This
 also has a large effect on the $\eta$ parameter as the addition
 breaks the remaining correlations and allows the CMB and weak lensing
 observations to constrain $\eta$. The effect of adding Stage I
 cluster counts is shown in Figure~\ref{fig:tjenk1} As mentioned
 earlier, there are some uncertainties associated with cluster
 counts. As can be seen readily from Figure~\ref{fig:tjenk1}, although
 marginalising over these uncertainties reduces the impact of
 clusters, they still add to the constraining power of the CMB and
 Weak lensing.

The comparison for Stage I, II, and III Fisher results in $\eta$ and
$\mu$ are shown in Figure~\ref{fig:threestage}. The effect of adding a
marginalisation over the mass assignment uncertainty is shown in
Figure~\ref{fig:jenk3}. The constraints worsen by a factor of between 2
and 3 when the uncertainty is taken into account but we stress that
our estimates are conservative since we have not allowed for any
improvement in $\sigma_{\cal M}$ in successive Stages.

A summary of the results is shown in Table~\ref{table:jenk1} for all
three Stages. The forecasted errors $\Delta \eta$  and $\Delta \mu$
are shown with and without addition of cluster counts (CC). All
constraints use the cross--correlation signal between CMB and weak
lensing (CMB $\otimes$ WL) and the values in brackets show the effect
of adding in the mass assignment uncertainty to the cluster counts. 

In general, we find that the MGPs can be constrained to within a few
percent using our forecasted surveys. The $\mu$ parameters is most
constrained by the observations. This is not surprising since it is
the parameter that affects the growth history directly. Including
the mass assignment error in the cluster counts reduces the
constraints by approximately a factor of 2. The improvement obtained
from successive Stages is limited, however the Stages recover the
constraining power that was lost to the mass assignment error of the
cluster counts. This reinforces the need for increased sensitivity and
survey size for future observations as it hedges against systematic
effects such as mass and photometric redshift resolution taken into
account here. 

To gauge the robustness of our predictions with respect to our choice
of fixed $\eta$ and $\mu$, we add a linear time dependence in the MGPs
of the form
\begin{equation}\label{eq:timedep}
 \mu(z)=(\mu_0-1)\left(\frac{z_{\rm mg}-z}{z_{\rm mg}} \right)+1 \, ,
\end{equation}
with a similar expression for $\eta$. Here, $\mu_0$ or $\eta_0$ is the
value of the MGP today ($z=0$), which is the parameter that will be
constrained in the Fisher matrix analysis. As such, the choice of
$z_{\rm mg}$, which is not included as a parameter in the Fisher
analysis, fixes the gradient of the linear time dependence. We are
therefore testing our results with respect to a choice of $z_{\rm mg}$
rather than extending the parameter space to include a parametrised
time dependence. 

As before, $z_{\rm mg}$ is the redshift at which the modification to
gravity `switches on'.  As shown in Figure~\ref{fig:td_jenk1},
allowing for this simple time dependence in the MGPs does not have a
large impact on constraints for the case where $z_{\rm
  mg}=30$. Although the time dependence reduces the impact of the MGPs
on the observations for the same value of the MGPs today, the relative
effect is also different in each of the signals and leads to a small
change in degeneracies. An additional signal is sourced in this case
since the time dependence of $\eta$ will affect the late--time
ISW. These combination of effects partly cancel, which results in a
small effect on the final constraints. It should be stressed that
there are many different choices of time dependence for the MGPs and
(\ref{eq:timedep}) is only one of these. In principle some other form
of time dependence could lead to much larger effects in this type of
comparison. A theoretically motivated for for the time dependence
would be better justified, however, this would build in a model
dependence which we are explicitly avoiding in this work.

Our results for $z_{\rm mg}=30$, 3, and 1 are summarised in
Table~\ref{table:timedep} which can be compared to the Stage I results
of Table~\ref{table:jenk1}. We find the effect becomes large only at
relatively low redshifts with the choice of $z_{\rm mg}=1$ leading to
an order of magnitude increase in Fisher errors. This result is not
indicative of the ability of the observations to constrain any possible time
behaviour of the modified gravity theory but rather it illustrates the
sensitivity of the observables to the departure from GR integrated
over time. 

\begin{table*}[t]
\centering
\begin{tabular}{|c| c| c|c|c|c|c|}
\hline
Stage I &\multicolumn{2}{|c|}{$z_{\rm mg}=30$}&\multicolumn{2}{|c|}{$z_{\rm 
mg}=3$}&\multicolumn{2}{|c|}{$z_{\rm mg}=1$}\\
\hline
Parameter & CMB $\otimes$ WL & CMB $\otimes$ WL + CC& CMB $\otimes$ WL & 
CMB $\otimes$ WL + CC& CMB $\otimes$ WL & CMB $\otimes$ WL + CC
\\
\hline
$\Delta\eta$ & $2.6 \times 10^{-2}$  & $6.4 \times 10^{-3} (1.5\times 
10^{-2})$ & $3.3 \times 10^{-2}$  & $1.9 \times 10^{-2} (2.6\times 
10^{-2})$ & $1.1 \times 10^{-1}$  & $1.1 \times 10^{-1} (1.1\times 
10^{-1})$ \\
$\Delta\mu$  & $9.1\times 10^{-3}$ & $3.5 \times 10^{-3}(5.6 \times 
10^{-3}) $& $1.0\times 10^{-2}$ & $5.0 \times 10^{-3}(7.6 \times 
10^{-3}) $& $5.1\times 10^{-2}$ & $5.1 \times 10^{-2}(5.1 \times 
10^{-2}) $\\[1ex]
\hline   
\end{tabular}
\caption{Fisher constraints for time--dependent MGPs for Stage I
   observations.}
\label{table:timedep} 
\end{table*}

\section{Discussion}\label{sec:disc}

Future generations of large area surveys hold much promise in testing
the validity of GR. In this work we have examined the impact, on MGP
constraints, of future cosmic shear and cluster count data used in
combination with CMB measurement. We have found that this combination
can constrain the MGPs to sub--percent accuracy. In particular the
inclusion of cluster counts, which are highly sensitive to any change
in the growth of the matter perturbations, adds a strong refinement in
the search for any deviations from the standard GR values of the
MGPs. However we have also shown that the inclusion of a simple model
for calibration uncertainties in the counts can affect the errors.

Our Fisher matrix analysis has shown that the MGPs may be
{\sl statistically} constrained to fractional levels of below a
percent. This clearly supercedes the current modeling accuracy of
weak lensing and cluster abundances in particular. In the case of weak
lensing, even at large scales, non-linear contributions to the matter
power spectrum are not modeled accurately enough whilst in the cluster
case the mass function formalism is certainly not accurate at the 1\%
level. Any analysis of actual data will necessarily have to include a
marginalisation over these and other modeling uncertainties. This can
be done either by including the model uncertainties in the final
Fisher matrix and marginalising by cutting out the directions in its
inverse or by including the model parameters as extra directions in
MCMC methods. Future MGP constraints may well be limited by such
systematic effects rather than the statistical limits of the observations.

There are, of course, other observables that can constrain the growth
of matter perturbations such as large scale structure, peculiar
velocities, redshift space distortions and future high redshift 21cm
surveys. These could all replace our choice of cluster counts as a
signal used to break the degeneracy between gravitational
potentials. All of these, however, present a number of problems either in
their interpretation as biased tracers or in the technological
challenges involved in large scale surveys. The observations of
clusters on the other hand is a relatively simple procedure and SZ
surveys of clusters are a necessary by--product of current and future
CMB experiments. The only difficulty involved in this signal is the
theoretical modelling required to use them to obtain constraints due
to the uncertainty in the choice of mass functions to be used for
theories of modified gravity.

In obtaining our forecasted constraints we have only employed the
simplest description of phenomenologically modified gravity. The model
building for such theories has already shown that there there is much
more complexity to explore. All models must incorporate a screening
mechanism to satisfy standard solar system and laboratory tests of
gravity. Screening mechanism are only one way in which modified
gravity models may include significant scale dependence, which we
have not taken into account in this work and may provide more testable
predictions and therefore more ability to differentiate between
models. Concrete models may also lead to modifications to the
background expansion, which we have not explored here. This can also
lead to additional constraints on model parameters.

Allowing for a very general scale dependence of the MGPs will inevitably 
introduce a scale dependence into the constraints. In particular our 
combination of data generates constraints based on large scales, due to 
the ISW effect, and a fairly narrow range of scales of around 10 Mpc 
from cluster counts. If there is a significant scale dependence of the 
MGPs between the two scales then the combination of data considered here 
would give stronger constraints on some scales than others. A complete 
treatment would include a Fisher eigenmode analysis for an extended 
range of parameters describing the scale dependence of the MGPs \cite{gbz10,gbz10b,gbz09}. 
We leave this for future work for the combination of data considered here.

The experimental outlook is promising for the observations we have
dealt with in this work. Over the next 5-10 years, deviations from GR
should be well constrained, and the concordance cosmology will either
be more secure or may even have undergone a paradigm shift. A null
result would support the concordance cosmology, a conclusion that
would be even stronger if dark matter had been detected
non-gravitationally by then. A detection of a deviation from GR would
be potentially more interesting but would require a completely new
theoretical framework and trigger a search for an underlying model for
the modifications.  Of course, a dark energy model with perturbations
may turn out to fit the data just as well and it is not clear at this
point whether this degeneracy will ever be broken by observations.

{\sl Acknowledgements} We thank Filipe Abdalla for pointing out the
work on self calibration of clusters to us. This work was supported by
an STFC studentship.

\appendix
\section{Redshift dependent limiting mass for cluster surveys}
\label{app:mlimz}
The full treatment of the limiting mass here is based on \cite{mlim}.
The limiting mass at redshift $z$, for a detector at dimensionless 
frequency $x$ with a flux limit of $S_{\rm lim}$ is given by
\begin{equation}
 M_{\rm lim}=\left(\frac{S_{\rm
       lim}d^2_A(z)}{f(x)f_{b}A_{\rm SZ}(1+z)}\right)^{\frac{1}{\beta_{\rm
       SZ}}}\,.
\end{equation}
Here, $d_A(z)$ is the angular diameter distance, $f_b$ is the baryon fraction
and $f(x)$ is the frequency dependence of the SZ effect, given by
$$f(x)=\frac{x^4e^x}{(e^x-1)^2}\left(\frac{x(e^x+1)}{e^x-1}-4\right)\,.$$
The dimensionless frequency is given by $x=\nu\times
\frac{h}{k_{B}T_{\rm CMB}}$,
with $T_{\rm CMB}$ the temperature of the CMB. For our analysis based on
Planck, we used $S_{\rm lim}=30$ mJy and a frequency $\nu=353$
GHz. Both $A_{\rm SZ}$
and $\beta_{\rm SZ}$ are parameters from the mass-temperature relation for
clusters. The values of the constants used here are as follows:
$f_{\rm ICM}=0.06$, $\beta_{\rm SZ}=1.75$, and $A_{\rm SZ}=3.781\times 10^8$.

\bibliography{mn}

\begin{thebibliography}{73}
\expandafter\ifx\csname natexlab\endcsname\relax\def\natexlab#1{#1}\fi
\expandafter\ifx\csname bibnamefont\endcsname\relax
  \def\bibnamefont#1{#1}\fi
\expandafter\ifx\csname bibfnamefont\endcsname\relax
  \def\bibfnamefont#1{#1}\fi
\expandafter\ifx\csname citenamefont\endcsname\relax
  \def\citenamefont#1{#1}\fi
\expandafter\ifx\csname url\endcsname\relax
  \def\url#1{\texttt{#1}}\fi
\expandafter\ifx\csname urlprefix\endcsname\relax\def\urlprefix{URL }\fi
\providecommand{\bibinfo}[2]{#2}
\providecommand{\eprint}[2][]{\url{#2}}

\bibitem[{\citenamefont{{Dvali} et~al.}(2000)\citenamefont{{Dvali},
  {Gabadadze}, and {Porrati}}}]{dgp}
\bibinfo{author}{\bibfnamefont{G.}~\bibnamefont{{Dvali}}},
  \bibinfo{author}{\bibfnamefont{G.}~\bibnamefont{{Gabadadze}}},
  \bibnamefont{and}
  \bibinfo{author}{\bibfnamefont{M.}~\bibnamefont{{Porrati}}},
  \bibinfo{journal}{Physics Letters B} \textbf{\bibinfo{volume}{485}},
  \bibinfo{pages}{208} (\bibinfo{year}{2000}), \eprint{arXiv:hep-th/0005016}.

\bibitem[{\citenamefont{{Mannheim}}(2001)}]{cfg}
\bibinfo{author}{\bibfnamefont{P.~D.} \bibnamefont{{Mannheim}}},
  \bibinfo{journal}{\apj} \textbf{\bibinfo{volume}{561}}, \bibinfo{pages}{1}
  (\bibinfo{year}{2001}), \eprint{arXiv:astro-ph/9910093}.

\bibitem[{\citenamefont{{Milgrom}}(1983)}]{mond}
\bibinfo{author}{\bibfnamefont{M.}~\bibnamefont{{Milgrom}}},
  \bibinfo{journal}{\apj} \textbf{\bibinfo{volume}{270}}, \bibinfo{pages}{365}
  (\bibinfo{year}{1983}).

\bibitem[{\citenamefont{{Sanders}}(2005)}]{teves}
\bibinfo{author}{\bibfnamefont{R.~H.} \bibnamefont{{Sanders}}},
  \bibinfo{journal}{\mnras} \textbf{\bibinfo{volume}{363}},
  \bibinfo{pages}{459} (\bibinfo{year}{2005}), \eprint{arXiv:astro-ph/0502222}.

\bibitem[{\citenamefont{{Jacobson} and {Mattingly}}(2001)}]{eatheories}
\bibinfo{author}{\bibfnamefont{T.}~\bibnamefont{{Jacobson}}} \bibnamefont{and}
  \bibinfo{author}{\bibfnamefont{D.}~\bibnamefont{{Mattingly}}},
  \bibinfo{journal}{\prd} \textbf{\bibinfo{volume}{64}},
  \bibinfo{pages}{024028} (\bibinfo{year}{2001}), \eprint{arXiv:gr-qc/0007031}.

\bibitem[{\citenamefont{{Clifton} et~al.}(2011)\citenamefont{{Clifton},
  {Ferreira}, {Padilla}, and {Skordis}}}]{mgreview}
\bibinfo{author}{\bibfnamefont{T.}~\bibnamefont{{Clifton}}},
  \bibinfo{author}{\bibfnamefont{P.~G.} \bibnamefont{{Ferreira}}},
  \bibinfo{author}{\bibfnamefont{A.}~\bibnamefont{{Padilla}}},
  \bibnamefont{and}
  \bibinfo{author}{\bibfnamefont{C.}~\bibnamefont{{Skordis}}},
  \bibinfo{journal}{ArXiv e-prints}  (\bibinfo{year}{2011}),
  \eprint{1106.2476}.

\bibitem[{\citenamefont{{Song} and {Percival}}(2009)}]{song09}
\bibinfo{author}{\bibfnamefont{Y.}~\bibnamefont{{Song}}} \bibnamefont{and}
  \bibinfo{author}{\bibfnamefont{W.~J.} \bibnamefont{{Percival}}},
  \bibinfo{journal}{\jcap} \textbf{\bibinfo{volume}{10}}, \bibinfo{pages}{4}
  (\bibinfo{year}{2009}), \eprint{0807.0810}.

\bibitem[{\citenamefont{{Shapiro} et~al.}(2010)\citenamefont{{Shapiro},
  {Dodelson}, {Hoyle}, {Samushia}, and {Flaugher}}}]{shapiro10}
\bibinfo{author}{\bibfnamefont{C.}~\bibnamefont{{Shapiro}}},
  \bibinfo{author}{\bibfnamefont{S.}~\bibnamefont{{Dodelson}}},
  \bibinfo{author}{\bibfnamefont{B.}~\bibnamefont{{Hoyle}}},
  \bibinfo{author}{\bibfnamefont{L.}~\bibnamefont{{Samushia}}},
  \bibnamefont{and}
  \bibinfo{author}{\bibfnamefont{B.}~\bibnamefont{{Flaugher}}},
  \bibinfo{journal}{\prd} \textbf{\bibinfo{volume}{82}},
  \bibinfo{pages}{043520} (\bibinfo{year}{2010}), \eprint{1004.4810}.

\bibitem[{\citenamefont{{Acquaviva} and {Gawiser}}(2010)}]{acqua10}
\bibinfo{author}{\bibfnamefont{V.}~\bibnamefont{{Acquaviva}}} \bibnamefont{and}
  \bibinfo{author}{\bibfnamefont{E.}~\bibnamefont{{Gawiser}}},
  \bibinfo{journal}{ArXiv e-prints}  (\bibinfo{year}{2010}),
  \eprint{1008.3392}.

\bibitem[{\citenamefont{{Linder}}(2005)}]{linder05}
\bibinfo{author}{\bibfnamefont{E.~V.} \bibnamefont{{Linder}}},
  \bibinfo{journal}{\prd} \textbf{\bibinfo{volume}{72}},
  \bibinfo{pages}{043529} (\bibinfo{year}{2005}),
  \eprint{arXiv:astro-ph/0507263}.

\bibitem[{\citenamefont{{Wang} and {Steinhardt}}(1998)}]{wang98}
\bibinfo{author}{\bibfnamefont{L.}~\bibnamefont{{Wang}}} \bibnamefont{and}
  \bibinfo{author}{\bibfnamefont{P.~J.} \bibnamefont{{Steinhardt}}},
  \bibinfo{journal}{\apj} \textbf{\bibinfo{volume}{508}}, \bibinfo{pages}{483}
  (\bibinfo{year}{1998}), \eprint{arXiv:astro-ph/9804015}.

\bibitem[{\citenamefont{{Pogosian} et~al.}(2010)\citenamefont{{Pogosian},
  {Silvestri}, {Koyama}, and {Zhao}}}]{gbz10b}
\bibinfo{author}{\bibfnamefont{L.}~\bibnamefont{{Pogosian}}},
  \bibinfo{author}{\bibfnamefont{A.}~\bibnamefont{{Silvestri}}},
  \bibinfo{author}{\bibfnamefont{K.}~\bibnamefont{{Koyama}}}, \bibnamefont{and}
  \bibinfo{author}{\bibfnamefont{G.}~\bibnamefont{{Zhao}}},
  \bibinfo{journal}{\prd} \textbf{\bibinfo{volume}{81}},
  \bibinfo{pages}{104023} (\bibinfo{year}{2010}), \eprint{1002.2382}.

\bibitem[{\citenamefont{{Zhao} et~al.}(2009{\natexlab{a}})\citenamefont{{Zhao},
  {Pogosian}, {Silvestri}, and {Zylberberg}}}]{gbz09}
\bibinfo{author}{\bibfnamefont{G.}~\bibnamefont{{Zhao}}},
  \bibinfo{author}{\bibfnamefont{L.}~\bibnamefont{{Pogosian}}},
  \bibinfo{author}{\bibfnamefont{A.}~\bibnamefont{{Silvestri}}},
  \bibnamefont{and}
  \bibinfo{author}{\bibfnamefont{J.}~\bibnamefont{{Zylberberg}}},
  \bibinfo{journal}{Physical Review Letters} \textbf{\bibinfo{volume}{103}},
  \bibinfo{pages}{241301} (\bibinfo{year}{2009}{\natexlab{a}}),
  \eprint{0905.1326}.

\bibitem[{\citenamefont{{Dossett} et~al.}(2010)\citenamefont{{Dossett},
  {Ishak}, {Moldenhauer}, {Gong}, and {Wang}}}]{dossett10}
\bibinfo{author}{\bibfnamefont{J.}~\bibnamefont{{Dossett}}},
  \bibinfo{author}{\bibfnamefont{M.}~\bibnamefont{{Ishak}}},
  \bibinfo{author}{\bibfnamefont{J.}~\bibnamefont{{Moldenhauer}}},
  \bibinfo{author}{\bibfnamefont{Y.}~\bibnamefont{{Gong}}}, \bibnamefont{and}
  \bibinfo{author}{\bibfnamefont{A.}~\bibnamefont{{Wang}}},
  \bibinfo{journal}{\jcap} \textbf{\bibinfo{volume}{4}}, \bibinfo{pages}{22}
  (\bibinfo{year}{2010}), \eprint{1004.3086}.

\bibitem[{\citenamefont{{Daniel} and {Linder}}(2010)}]{daniel10b}
\bibinfo{author}{\bibfnamefont{S.~F.} \bibnamefont{{Daniel}}} \bibnamefont{and}
  \bibinfo{author}{\bibfnamefont{E.~V.} \bibnamefont{{Linder}}},
  \bibinfo{journal}{ArXiv e-prints}  (\bibinfo{year}{2010}),
  \eprint{1008.0397}.

\bibitem[{\citenamefont{{Serra} et~al.}(2009)\citenamefont{{Serra}, {Cooray},
  {Daniel}, {Caldwell}, and {Melchiorri}}}]{serra09}
\bibinfo{author}{\bibfnamefont{P.}~\bibnamefont{{Serra}}},
  \bibinfo{author}{\bibfnamefont{A.}~\bibnamefont{{Cooray}}},
  \bibinfo{author}{\bibfnamefont{S.~F.} \bibnamefont{{Daniel}}},
  \bibinfo{author}{\bibfnamefont{R.}~\bibnamefont{{Caldwell}}},
  \bibnamefont{and}
  \bibinfo{author}{\bibfnamefont{A.}~\bibnamefont{{Melchiorri}}},
  \bibinfo{journal}{\prd} \textbf{\bibinfo{volume}{79}},
  \bibinfo{pages}{101301} (\bibinfo{year}{2009}), \eprint{0901.0917}.

\bibitem[{\citenamefont{{Guzik} et~al.}(2010)\citenamefont{{Guzik}, {Jain}, and
  {Takada}}}]{guzik10}
\bibinfo{author}{\bibfnamefont{J.}~\bibnamefont{{Guzik}}},
  \bibinfo{author}{\bibfnamefont{B.}~\bibnamefont{{Jain}}}, \bibnamefont{and}
  \bibinfo{author}{\bibfnamefont{M.}~\bibnamefont{{Takada}}},
  \bibinfo{journal}{\prd} \textbf{\bibinfo{volume}{81}},
  \bibinfo{pages}{023503} (\bibinfo{year}{2010}), \eprint{0906.2221}.

\bibitem[{\citenamefont{{Kosowsky} and {Bhattacharya}}(2009)}]{kosow09}
\bibinfo{author}{\bibfnamefont{A.}~\bibnamefont{{Kosowsky}}} \bibnamefont{and}
  \bibinfo{author}{\bibfnamefont{S.}~\bibnamefont{{Bhattacharya}}},
  \bibinfo{journal}{\prd} \textbf{\bibinfo{volume}{80}},
  \bibinfo{pages}{062003} (\bibinfo{year}{2009}), \eprint{0907.4202}.

\bibitem[{\citenamefont{{Heavens} et~al.}(2007)\citenamefont{{Heavens},
  {Kitching}, and {Verde}}}]{heavens07}
\bibinfo{author}{\bibfnamefont{A.~F.} \bibnamefont{{Heavens}}},
  \bibinfo{author}{\bibfnamefont{T.~D.} \bibnamefont{{Kitching}}},
  \bibnamefont{and} \bibinfo{author}{\bibfnamefont{L.}~\bibnamefont{{Verde}}},
  \bibinfo{journal}{\mnras} \textbf{\bibinfo{volume}{380}},
  \bibinfo{pages}{1029} (\bibinfo{year}{2007}),
  \eprint{arXiv:astro-ph/0703191}.

\bibitem[{\citenamefont{{The Planck Collaboration}}(2006)}]{planck}
\bibinfo{author}{\bibnamefont{{The Planck Collaboration}}},
  \bibinfo{journal}{ArXiv Astrophysics e-prints}  (\bibinfo{year}{2006}),
  \eprint{arXiv:astro-ph/0604069}.

\bibitem[{\citenamefont{{Hoekstra} et~al.}(2006)\citenamefont{{Hoekstra},
  {Mellier}, {van Waerbeke}, {Semboloni}, {Fu}, {Hudson}, {Parker}, {Tereno},
  and {Benabed}}}]{cfhtls}
\bibinfo{author}{\bibfnamefont{H.}~\bibnamefont{{Hoekstra}}},
  \bibinfo{author}{\bibfnamefont{Y.}~\bibnamefont{{Mellier}}},
  \bibinfo{author}{\bibfnamefont{L.}~\bibnamefont{{van Waerbeke}}},
  \bibinfo{author}{\bibfnamefont{E.}~\bibnamefont{{Semboloni}}},
  \bibinfo{author}{\bibfnamefont{L.}~\bibnamefont{{Fu}}},
  \bibinfo{author}{\bibfnamefont{M.~J.} \bibnamefont{{Hudson}}},
  \bibinfo{author}{\bibfnamefont{L.~C.} \bibnamefont{{Parker}}},
  \bibinfo{author}{\bibfnamefont{I.}~\bibnamefont{{Tereno}}}, \bibnamefont{and}
  \bibinfo{author}{\bibfnamefont{K.}~\bibnamefont{{Benabed}}},
  \bibinfo{journal}{\apj} \textbf{\bibinfo{volume}{647}}, \bibinfo{pages}{116}
  (\bibinfo{year}{2006}), \eprint{arXiv:astro-ph/0511089}.

\bibitem[{\citenamefont{{Kaiser}}(2004)}]{panstarrs}
\bibinfo{author}{\bibfnamefont{N.}~\bibnamefont{{Kaiser}}},
  \bibinfo{journal}{Society of Photo-Optical Instrumentation Engineers (SPIE)
  Conference Series} \textbf{\bibinfo{volume}{5489}}, \bibinfo{pages}{11}
  (\bibinfo{year}{2004}).

\bibitem[{\citenamefont{{The Dark Energy Survey Collaboration}}(2005)}]{des}
\bibinfo{author}{\bibnamefont{{The Dark Energy Survey Collaboration}}},
  \bibinfo{journal}{ArXiv Astrophysics e-prints}  (\bibinfo{year}{2005}),
  \eprint{arXiv:astro-ph/0510346}.

\bibitem[{\citenamefont{{Tyson}}(2002)}]{lsst}
\bibinfo{author}{\bibfnamefont{J.~A.} \bibnamefont{{Tyson}}},
  \bibinfo{journal}{Society of Photo-Optical Instrumentation Engineers (SPIE)
  Conference Series} \textbf{\bibinfo{volume}{4836}}, \bibinfo{pages}{10}
  (\bibinfo{year}{2002}).

\bibitem[{\citenamefont{{Aihara} et~al.}(2011)\citenamefont{{Aihara}, {Allende
  Prieto}, {An}, {Anderson}, {Aubourg}, {Balbinot}, {Beers}, {Berlind},
  {Bickerton}, {Bizyaev} et~al.}}]{sdss}
\bibinfo{author}{\bibfnamefont{H.}~\bibnamefont{{Aihara}}},
  \bibinfo{author}{\bibfnamefont{C.}~\bibnamefont{{Allende Prieto}}},
  \bibinfo{author}{\bibfnamefont{D.}~\bibnamefont{{An}}},
  \bibinfo{author}{\bibfnamefont{S.~F.} \bibnamefont{{Anderson}}},
  \bibinfo{author}{\bibfnamefont{{\'E}.}~\bibnamefont{{Aubourg}}},
  \bibinfo{author}{\bibfnamefont{E.}~\bibnamefont{{Balbinot}}},
  \bibinfo{author}{\bibfnamefont{T.~C.} \bibnamefont{{Beers}}},
  \bibinfo{author}{\bibfnamefont{A.~A.} \bibnamefont{{Berlind}}},
  \bibinfo{author}{\bibfnamefont{S.~J.} \bibnamefont{{Bickerton}}},
  \bibinfo{author}{\bibfnamefont{D.}~\bibnamefont{{Bizyaev}}},
  \bibnamefont{et~al.}, \bibinfo{journal}{\apjs}
  \textbf{\bibinfo{volume}{193}}, \bibinfo{pages}{29} (\bibinfo{year}{2011}),
  \eprint{1101.1559}.

\bibitem[{\citenamefont{{Colless}}(1999)}]{2dfgrs}
\bibinfo{author}{\bibfnamefont{M.}~\bibnamefont{{Colless}}},
  \bibinfo{journal}{Royal Society of London Philosophical Transactions Series
  A} \textbf{\bibinfo{volume}{357}}, \bibinfo{pages}{105}
  (\bibinfo{year}{1999}), \eprint{arXiv:astro-ph/9804079}.

\bibitem[{\citenamefont{{Bertschinger} and {Zukin}}(2008)}]{bert08}
\bibinfo{author}{\bibfnamefont{E.}~\bibnamefont{{Bertschinger}}}
  \bibnamefont{and} \bibinfo{author}{\bibfnamefont{P.}~\bibnamefont{{Zukin}}},
  \bibinfo{journal}{\prd} \textbf{\bibinfo{volume}{78}},
  \bibinfo{pages}{024015} (\bibinfo{year}{2008}), \eprint{0801.2431}.

\bibitem[{\citenamefont{{Kunz} and {Sapone}}(2007)}]{kunz06}
\bibinfo{author}{\bibfnamefont{M.}~\bibnamefont{{Kunz}}} \bibnamefont{and}
  \bibinfo{author}{\bibfnamefont{D.}~\bibnamefont{{Sapone}}},
  \bibinfo{journal}{Physical Review Letters} \textbf{\bibinfo{volume}{98}},
  \bibinfo{pages}{121301} (\bibinfo{year}{2007}),
  \eprint{arXiv:astro-ph/0612452}.

\bibitem[{\citenamefont{{Daniel} et~al.}(2010)\citenamefont{{Daniel}, {Linder},
  {Smith}, {Caldwell}, {Cooray}, {Leauthaud}, and {Lombriser}}}]{daniel10}
\bibinfo{author}{\bibfnamefont{S.~F.} \bibnamefont{{Daniel}}},
  \bibinfo{author}{\bibfnamefont{E.~V.} \bibnamefont{{Linder}}},
  \bibinfo{author}{\bibfnamefont{T.~L.} \bibnamefont{{Smith}}},
  \bibinfo{author}{\bibfnamefont{R.~R.} \bibnamefont{{Caldwell}}},
  \bibinfo{author}{\bibfnamefont{A.}~\bibnamefont{{Cooray}}},
  \bibinfo{author}{\bibfnamefont{A.}~\bibnamefont{{Leauthaud}}},
  \bibnamefont{and}
  \bibinfo{author}{\bibfnamefont{L.}~\bibnamefont{{Lombriser}}},
  \bibinfo{journal}{\prd} \textbf{\bibinfo{volume}{81}},
  \bibinfo{pages}{123508} (\bibinfo{year}{2010}), \eprint{1002.1962}.

\bibitem[{\citenamefont{{Zhao} et~al.}(2009{\natexlab{b}})\citenamefont{{Zhao},
  {Pogosian}, {Silvestri}, and {Zylberberg}}}]{mgcamb}
\bibinfo{author}{\bibfnamefont{G.}~\bibnamefont{{Zhao}}},
  \bibinfo{author}{\bibfnamefont{L.}~\bibnamefont{{Pogosian}}},
  \bibinfo{author}{\bibfnamefont{A.}~\bibnamefont{{Silvestri}}},
  \bibnamefont{and}
  \bibinfo{author}{\bibfnamefont{J.}~\bibnamefont{{Zylberberg}}},
  \bibinfo{journal}{\prd} \textbf{\bibinfo{volume}{79}},
  \bibinfo{pages}{083513} (\bibinfo{year}{2009}{\natexlab{b}}),
  \eprint{0809.3791}.

\bibitem[{\citenamefont{{Lewis} et~al.}(2000)\citenamefont{{Lewis},
  {Challinor}, and {Lasenby}}}]{camb}
\bibinfo{author}{\bibfnamefont{A.}~\bibnamefont{{Lewis}}},
  \bibinfo{author}{\bibfnamefont{A.}~\bibnamefont{{Challinor}}},
  \bibnamefont{and}
  \bibinfo{author}{\bibfnamefont{A.}~\bibnamefont{{Lasenby}}},
  \bibinfo{journal}{\apj} \textbf{\bibinfo{volume}{538}}, \bibinfo{pages}{473}
  (\bibinfo{year}{2000}), \eprint{arXiv:astro-ph/9911177}.

\bibitem[{\citenamefont{{Skordis}}(2009)}]{skordis09}
\bibinfo{author}{\bibfnamefont{C.}~\bibnamefont{{Skordis}}},
  \bibinfo{journal}{\prd} \textbf{\bibinfo{volume}{79}},
  \bibinfo{pages}{123527} (\bibinfo{year}{2009}), \eprint{0806.1238}.

\bibitem[{\citenamefont{{Contaldi} and {Thomas}}({in preparation})}]{inprep}
\bibinfo{author}{\bibfnamefont{C.}~\bibnamefont{{Contaldi}}} \bibnamefont{and}
  \bibinfo{author}{\bibfnamefont{D.~B.} \bibnamefont{{Thomas}}}
  (\bibinfo{year}{{in preparation}}).

\bibitem[{\citenamefont{{Jarosik} et~al.}(2011)\citenamefont{{Jarosik},
  {Bennett}, {Dunkley}, {Gold}, {Greason}, {Halpern}, {Hill}, {Hinshaw},
  {Kogut}, {Komatsu} et~al.}}]{wmap}
\bibinfo{author}{\bibfnamefont{N.}~\bibnamefont{{Jarosik}}},
  \bibinfo{author}{\bibfnamefont{C.~L.} \bibnamefont{{Bennett}}},
  \bibinfo{author}{\bibfnamefont{J.}~\bibnamefont{{Dunkley}}},
  \bibinfo{author}{\bibfnamefont{B.}~\bibnamefont{{Gold}}},
  \bibinfo{author}{\bibfnamefont{M.~R.} \bibnamefont{{Greason}}},
  \bibinfo{author}{\bibfnamefont{M.}~\bibnamefont{{Halpern}}},
  \bibinfo{author}{\bibfnamefont{R.~S.} \bibnamefont{{Hill}}},
  \bibinfo{author}{\bibfnamefont{G.}~\bibnamefont{{Hinshaw}}},
  \bibinfo{author}{\bibfnamefont{A.}~\bibnamefont{{Kogut}}},
  \bibinfo{author}{\bibfnamefont{E.}~\bibnamefont{{Komatsu}}},
  \bibnamefont{et~al.}, \bibinfo{journal}{\apjs}
  \textbf{\bibinfo{volume}{192}}, \bibinfo{pages}{14} (\bibinfo{year}{2011}),
  \eprint{1001.4744}.

\bibitem[{\citenamefont{{Manera} and {Mota}}(2006)}]{manera06}
\bibinfo{author}{\bibfnamefont{M.}~\bibnamefont{{Manera}}} \bibnamefont{and}
  \bibinfo{author}{\bibfnamefont{D.~F.} \bibnamefont{{Mota}}},
  \bibinfo{journal}{\mnras} \textbf{\bibinfo{volume}{371}},
  \bibinfo{pages}{1373} (\bibinfo{year}{2006}),
  \eprint{arXiv:astro-ph/0504519}.

\bibitem[{\citenamefont{{Waizmann} and {Bartelmann}}(2009)}]{waiz10}
\bibinfo{author}{\bibfnamefont{J.}~\bibnamefont{{Waizmann}}} \bibnamefont{and}
  \bibinfo{author}{\bibfnamefont{M.}~\bibnamefont{{Bartelmann}}},
  \bibinfo{journal}{\aap} \textbf{\bibinfo{volume}{493}}, \bibinfo{pages}{859}
  (\bibinfo{year}{2009}), \eprint{0804.2815}.

\bibitem[{\citenamefont{{Alam} et~al.}(2010)\citenamefont{{Alam}, {Luki{\'c}},
  and {Bhattacharya}}}]{alam10}
\bibinfo{author}{\bibfnamefont{U.}~\bibnamefont{{Alam}}},
  \bibinfo{author}{\bibfnamefont{Z.}~\bibnamefont{{Luki{\'c}}}},
  \bibnamefont{and}
  \bibinfo{author}{\bibfnamefont{S.}~\bibnamefont{{Bhattacharya}}},
  \bibinfo{journal}{ArXiv e-prints}  (\bibinfo{year}{2010}),
  \eprint{1004.0437}.

\bibitem[{\citenamefont{{Erlich} et~al.}(2008)\citenamefont{{Erlich}, {Glover},
  and {Weiner}}}]{erlich08}
\bibinfo{author}{\bibfnamefont{J.}~\bibnamefont{{Erlich}}},
  \bibinfo{author}{\bibfnamefont{B.}~\bibnamefont{{Glover}}}, \bibnamefont{and}
  \bibinfo{author}{\bibfnamefont{N.}~\bibnamefont{{Weiner}}},
  \bibinfo{journal}{\jcap} \textbf{\bibinfo{volume}{3}}, \bibinfo{pages}{6}
  (\bibinfo{year}{2008}), \eprint{0709.3442}.

\bibitem[{\citenamefont{{Battye} and {Weller}}(2003)}]{battye03}
\bibinfo{author}{\bibfnamefont{R.~A.} \bibnamefont{{Battye}}} \bibnamefont{and}
  \bibinfo{author}{\bibfnamefont{J.}~\bibnamefont{{Weller}}},
  \bibinfo{journal}{\prd} \textbf{\bibinfo{volume}{68}},
  \bibinfo{pages}{083506} (\bibinfo{year}{2003}),
  \eprint{arXiv:astro-ph/0305568}.

\bibitem[{\citenamefont{{Basilakos} et~al.}(2010)\citenamefont{{Basilakos},
  {Plionis}, and {Lima}}}]{basil10}
\bibinfo{author}{\bibfnamefont{S.}~\bibnamefont{{Basilakos}}},
  \bibinfo{author}{\bibfnamefont{M.}~\bibnamefont{{Plionis}}},
  \bibnamefont{and} \bibinfo{author}{\bibfnamefont{J.~A.~S.}
  \bibnamefont{{Lima}}}, \bibinfo{journal}{ArXiv e-prints}
  (\bibinfo{year}{2010}), \eprint{1006.3418}.

\bibitem[{\citenamefont{{Rapetti} et~al.}(2010)\citenamefont{{Rapetti},
  {Allen}, {Mantz}, and {Ebeling}}}]{rapetti10}
\bibinfo{author}{\bibfnamefont{D.}~\bibnamefont{{Rapetti}}},
  \bibinfo{author}{\bibfnamefont{S.~W.} \bibnamefont{{Allen}}},
  \bibinfo{author}{\bibfnamefont{A.}~\bibnamefont{{Mantz}}}, \bibnamefont{and}
  \bibinfo{author}{\bibfnamefont{H.}~\bibnamefont{{Ebeling}}},
  \bibinfo{journal}{\mnras} \textbf{\bibinfo{volume}{406}},
  \bibinfo{pages}{1796} (\bibinfo{year}{2010}), \eprint{0911.1787}.

\bibitem[{\citenamefont{{Tang} et~al.}(2006)\citenamefont{{Tang}, {Weller}, and
  {Zablocki}}}]{tang06}
\bibinfo{author}{\bibfnamefont{J.}~\bibnamefont{{Tang}}},
  \bibinfo{author}{\bibfnamefont{J.}~\bibnamefont{{Weller}}}, \bibnamefont{and}
  \bibinfo{author}{\bibfnamefont{A.}~\bibnamefont{{Zablocki}}},
  \bibinfo{journal}{ArXiv Astrophysics e-prints}  (\bibinfo{year}{2006}),
  \eprint{arXiv:astro-ph/0609028}.

\bibitem[{\citenamefont{{Schmidt}}(2009{\natexlab{a}})}]{schmidt09}
\bibinfo{author}{\bibfnamefont{F.}~\bibnamefont{{Schmidt}}},
  \bibinfo{journal}{\prd} \textbf{\bibinfo{volume}{80}},
  \bibinfo{pages}{043001} (\bibinfo{year}{2009}{\natexlab{a}}),
  \eprint{0905.0858}.

\bibitem[{\citenamefont{{Kobayashi} and {Tashiro}}(2009)}]{kobay10}
\bibinfo{author}{\bibfnamefont{T.}~\bibnamefont{{Kobayashi}}} \bibnamefont{and}
  \bibinfo{author}{\bibfnamefont{H.}~\bibnamefont{{Tashiro}}},
  \bibinfo{journal}{\mnras} \textbf{\bibinfo{volume}{398}},
  \bibinfo{pages}{477} (\bibinfo{year}{2009}), \eprint{0903.3738}.

\bibitem[{\citenamefont{{Zhao} et~al.}(2010{\natexlab{a}})\citenamefont{{Zhao},
  {Macci{\`o}}, {Li}, {Hoekstra}, and {Feix}}}]{zhao10}
\bibinfo{author}{\bibfnamefont{H.}~\bibnamefont{{Zhao}}},
  \bibinfo{author}{\bibfnamefont{A.~V.} \bibnamefont{{Macci{\`o}}}},
  \bibinfo{author}{\bibfnamefont{B.}~\bibnamefont{{Li}}},
  \bibinfo{author}{\bibfnamefont{H.}~\bibnamefont{{Hoekstra}}},
  \bibnamefont{and} \bibinfo{author}{\bibfnamefont{M.}~\bibnamefont{{Feix}}},
  \bibinfo{journal}{\apjl} \textbf{\bibinfo{volume}{712}},
  \bibinfo{pages}{L179} (\bibinfo{year}{2010}{\natexlab{a}}),
  \eprint{0910.3207}.

\bibitem[{\citenamefont{{Li} et~al.}(2010)\citenamefont{{Li}, {Mota}, and
  {Barrow}}}]{li10}
\bibinfo{author}{\bibfnamefont{B.}~\bibnamefont{{Li}}},
  \bibinfo{author}{\bibfnamefont{D.~F.} \bibnamefont{{Mota}}},
  \bibnamefont{and} \bibinfo{author}{\bibfnamefont{J.~D.}
  \bibnamefont{{Barrow}}}, \bibinfo{journal}{ArXiv e-prints}
  (\bibinfo{year}{2010}), \eprint{1009.1400}.

\bibitem[{\citenamefont{{Schmidt} et~al.}(2009)\citenamefont{{Schmidt},
  {Vikhlinin}, and {Hu}}}]{schmidt09c}
\bibinfo{author}{\bibfnamefont{F.}~\bibnamefont{{Schmidt}}},
  \bibinfo{author}{\bibfnamefont{A.}~\bibnamefont{{Vikhlinin}}},
  \bibnamefont{and} \bibinfo{author}{\bibfnamefont{W.}~\bibnamefont{{Hu}}},
  \bibinfo{journal}{\prd} \textbf{\bibinfo{volume}{80}},
  \bibinfo{pages}{083505} (\bibinfo{year}{2009}), \eprint{0908.2457}.

\bibitem[{\citenamefont{{Khoury} and {Wyman}}(2009)}]{wyman09}
\bibinfo{author}{\bibfnamefont{J.}~\bibnamefont{{Khoury}}} \bibnamefont{and}
  \bibinfo{author}{\bibfnamefont{M.}~\bibnamefont{{Wyman}}},
  \bibinfo{journal}{\prd} \textbf{\bibinfo{volume}{80}},
  \bibinfo{pages}{064023} (\bibinfo{year}{2009}), \eprint{0903.1292}.

\bibitem[{\citenamefont{{Wyman} and {Khoury}}(2010)}]{wyman10}
\bibinfo{author}{\bibfnamefont{M.}~\bibnamefont{{Wyman}}} \bibnamefont{and}
  \bibinfo{author}{\bibfnamefont{J.}~\bibnamefont{{Khoury}}},
  \bibinfo{journal}{\prd} \textbf{\bibinfo{volume}{82}},
  \bibinfo{pages}{044032} (\bibinfo{year}{2010}), \eprint{1004.2046}.

\bibitem[{\citenamefont{{Schmidt}}(2009{\natexlab{b}})}]{schmidt09b}
\bibinfo{author}{\bibfnamefont{F.}~\bibnamefont{{Schmidt}}},
  \bibinfo{journal}{\prd} \textbf{\bibinfo{volume}{80}},
  \bibinfo{pages}{123003} (\bibinfo{year}{2009}{\natexlab{b}}),
  \eprint{0910.0235}.

\bibitem[{\citenamefont{{Macci{\`o}} et~al.}(2004)\citenamefont{{Macci{\`o}},
  {Quercellini}, {Mainini}, {Amendola}, and {Bonometto}}}]{macci04}
\bibinfo{author}{\bibfnamefont{A.~V.} \bibnamefont{{Macci{\`o}}}},
  \bibinfo{author}{\bibfnamefont{C.}~\bibnamefont{{Quercellini}}},
  \bibinfo{author}{\bibfnamefont{R.}~\bibnamefont{{Mainini}}},
  \bibinfo{author}{\bibfnamefont{L.}~\bibnamefont{{Amendola}}},
  \bibnamefont{and} \bibinfo{author}{\bibfnamefont{S.~A.}
  \bibnamefont{{Bonometto}}}, \bibinfo{journal}{\prd}
  \textbf{\bibinfo{volume}{69}}, \bibinfo{pages}{123516}
  (\bibinfo{year}{2004}), \eprint{arXiv:astro-ph/0309671}.

\bibitem[{\citenamefont{{Jenkins} et~al.}(2001)\citenamefont{{Jenkins},
  {Frenk}, {White}, {Colberg}, {Cole}, {Evrard}, {Couchman}, and
  {Yoshida}}}]{jenkins01}
\bibinfo{author}{\bibfnamefont{A.}~\bibnamefont{{Jenkins}}},
  \bibinfo{author}{\bibfnamefont{C.~S.} \bibnamefont{{Frenk}}},
  \bibinfo{author}{\bibfnamefont{S.~D.~M.} \bibnamefont{{White}}},
  \bibinfo{author}{\bibfnamefont{J.~M.} \bibnamefont{{Colberg}}},
  \bibinfo{author}{\bibfnamefont{S.}~\bibnamefont{{Cole}}},
  \bibinfo{author}{\bibfnamefont{A.~E.} \bibnamefont{{Evrard}}},
  \bibinfo{author}{\bibfnamefont{H.~M.~P.} \bibnamefont{{Couchman}}},
  \bibnamefont{and}
  \bibinfo{author}{\bibfnamefont{N.}~\bibnamefont{{Yoshida}}},
  \bibinfo{journal}{\mnras} \textbf{\bibinfo{volume}{321}},
  \bibinfo{pages}{372} (\bibinfo{year}{2001}), \eprint{arXiv:astro-ph/0005260}.

\bibitem[{\citenamefont{{Press} and {Schechter}}(1974)}]{ps}
\bibinfo{author}{\bibfnamefont{W.~H.} \bibnamefont{{Press}}} \bibnamefont{and}
  \bibinfo{author}{\bibfnamefont{P.}~\bibnamefont{{Schechter}}},
  \bibinfo{journal}{\apj} \textbf{\bibinfo{volume}{187}}, \bibinfo{pages}{425}
  (\bibinfo{year}{1974}).

\bibitem[{\citenamefont{{Bond} et~al.}(1991)\citenamefont{{Bond}, {Cole},
  {Efstathiou}, and {Kaiser}}}]{bond91}
\bibinfo{author}{\bibfnamefont{J.~R.} \bibnamefont{{Bond}}},
  \bibinfo{author}{\bibfnamefont{S.}~\bibnamefont{{Cole}}},
  \bibinfo{author}{\bibfnamefont{G.}~\bibnamefont{{Efstathiou}}},
  \bibnamefont{and} \bibinfo{author}{\bibfnamefont{N.}~\bibnamefont{{Kaiser}}},
  \bibinfo{journal}{\apj} \textbf{\bibinfo{volume}{379}}, \bibinfo{pages}{440}
  (\bibinfo{year}{1991}).

\bibitem[{\citenamefont{{Sheth} et~al.}(2001)\citenamefont{{Sheth}, {Mo}, and
  {Tormen}}}]{st99a}
\bibinfo{author}{\bibfnamefont{R.~K.} \bibnamefont{{Sheth}}},
  \bibinfo{author}{\bibfnamefont{H.~J.} \bibnamefont{{Mo}}}, \bibnamefont{and}
  \bibinfo{author}{\bibfnamefont{G.}~\bibnamefont{{Tormen}}},
  \bibinfo{journal}{\mnras} \textbf{\bibinfo{volume}{323}}, \bibinfo{pages}{1}
  (\bibinfo{year}{2001}), \eprint{arXiv:astro-ph/9907024}.

\bibitem[{\citenamefont{{Sheth} and {Tormen}}(1999)}]{st99b}
\bibinfo{author}{\bibfnamefont{R.~K.} \bibnamefont{{Sheth}}} \bibnamefont{and}
  \bibinfo{author}{\bibfnamefont{G.}~\bibnamefont{{Tormen}}},
  \bibinfo{journal}{\mnras} \textbf{\bibinfo{volume}{308}},
  \bibinfo{pages}{119} (\bibinfo{year}{1999}), \eprint{arXiv:astro-ph/9901122}.

\bibitem[{\citenamefont{{Sheth} and {Tormen}}(2002)}]{st02}
\bibinfo{author}{\bibfnamefont{R.~K.} \bibnamefont{{Sheth}}} \bibnamefont{and}
  \bibinfo{author}{\bibfnamefont{G.}~\bibnamefont{{Tormen}}},
  \bibinfo{journal}{\mnras} \textbf{\bibinfo{volume}{329}}, \bibinfo{pages}{61}
  (\bibinfo{year}{2002}), \eprint{arXiv:astro-ph/0105113}.

\bibitem[{\citenamefont{{Sunyaev} and {Zeldovich}}(1970)}]{sz1}
\bibinfo{author}{\bibfnamefont{R.~A.} \bibnamefont{{Sunyaev}}}
  \bibnamefont{and} \bibinfo{author}{\bibfnamefont{Y.~B.}
  \bibnamefont{{Zeldovich}}}, \bibinfo{journal}{Comments on Astrophysics and
  Space Physics} \textbf{\bibinfo{volume}{2}}, \bibinfo{pages}{66}
  (\bibinfo{year}{1970}).

\bibitem[{\citenamefont{{Norman}}(2010)}]{norman10}
\bibinfo{author}{\bibfnamefont{M.~L.} \bibnamefont{{Norman}}},
  \bibinfo{journal}{ArXiv e-prints}  (\bibinfo{year}{2010}),
  \eprint{1005.1100}.

\bibitem[{\citenamefont{{Lima} and {Hu}}(2005)}]{limahu05}
\bibinfo{author}{\bibfnamefont{M.}~\bibnamefont{{Lima}}} \bibnamefont{and}
  \bibinfo{author}{\bibfnamefont{W.}~\bibnamefont{{Hu}}},
  \bibinfo{journal}{\prd} \textbf{\bibinfo{volume}{72}},
  \bibinfo{pages}{043006} (\bibinfo{year}{2005}),
  \eprint{arXiv:astro-ph/0503363}.

\bibitem[{\citenamefont{{Crittenden} and {Turok}}(1996)}]{crit96}
\bibinfo{author}{\bibfnamefont{R.~G.} \bibnamefont{{Crittenden}}}
  \bibnamefont{and} \bibinfo{author}{\bibfnamefont{N.}~\bibnamefont{{Turok}}},
  \bibinfo{journal}{Physical Review Letters} \textbf{\bibinfo{volume}{76}},
  \bibinfo{pages}{575} (\bibinfo{year}{1996}), \eprint{arXiv:astro-ph/9510072}.

\bibitem[{\citenamefont{{Cooray}}(2002)}]{cooray02}
\bibinfo{author}{\bibfnamefont{A.}~\bibnamefont{{Cooray}}},
  \bibinfo{journal}{\prd} \textbf{\bibinfo{volume}{65}},
  \bibinfo{pages}{103510} (\bibinfo{year}{2002}),
  \eprint{arXiv:astro-ph/0112408}.

\bibitem[{\citenamefont{{Boughn} and {Crittenden}}(2004)}]{boughn04}
\bibinfo{author}{\bibfnamefont{S.}~\bibnamefont{{Boughn}}} \bibnamefont{and}
  \bibinfo{author}{\bibfnamefont{R.}~\bibnamefont{{Crittenden}}},
  \bibinfo{journal}{\nat} \textbf{\bibinfo{volume}{427}}, \bibinfo{pages}{45}
  (\bibinfo{year}{2004}), \eprint{arXiv:astro-ph/0305001}.

\bibitem[{\citenamefont{{Afshordi}}(2004)}]{afshord04}
\bibinfo{author}{\bibfnamefont{N.}~\bibnamefont{{Afshordi}}},
  \bibinfo{journal}{\prd} \textbf{\bibinfo{volume}{70}},
  \bibinfo{pages}{083536} (\bibinfo{year}{2004}),
  \eprint{arXiv:astro-ph/0401166}.

\bibitem[{\citenamefont{{Corasaniti} et~al.}(2005)\citenamefont{{Corasaniti},
  {Giannantonio}, and {Melchiorri}}}]{cora05}
\bibinfo{author}{\bibfnamefont{P.}~\bibnamefont{{Corasaniti}}},
  \bibinfo{author}{\bibfnamefont{T.}~\bibnamefont{{Giannantonio}}},
  \bibnamefont{and}
  \bibinfo{author}{\bibfnamefont{A.}~\bibnamefont{{Melchiorri}}},
  \bibinfo{journal}{\prd} \textbf{\bibinfo{volume}{71}},
  \bibinfo{pages}{123521} (\bibinfo{year}{2005}),
  \eprint{arXiv:astro-ph/0504115}.

\bibitem[{\citenamefont{{Giannantonio}
  et~al.}(2006)\citenamefont{{Giannantonio}, {Crittenden}, {Nichol},
  {Scranton}, {Richards}, {Myers}, {Brunner}, {Gray}, {Connolly}, and
  {Schneider}}}]{gian06}
\bibinfo{author}{\bibfnamefont{T.}~\bibnamefont{{Giannantonio}}},
  \bibinfo{author}{\bibfnamefont{R.~G.} \bibnamefont{{Crittenden}}},
  \bibinfo{author}{\bibfnamefont{R.~C.} \bibnamefont{{Nichol}}},
  \bibinfo{author}{\bibfnamefont{R.}~\bibnamefont{{Scranton}}},
  \bibinfo{author}{\bibfnamefont{G.~T.} \bibnamefont{{Richards}}},
  \bibinfo{author}{\bibfnamefont{A.~D.} \bibnamefont{{Myers}}},
  \bibinfo{author}{\bibfnamefont{R.~J.} \bibnamefont{{Brunner}}},
  \bibinfo{author}{\bibfnamefont{A.~G.} \bibnamefont{{Gray}}},
  \bibinfo{author}{\bibfnamefont{A.~J.} \bibnamefont{{Connolly}}},
  \bibnamefont{and} \bibinfo{author}{\bibfnamefont{D.~P.}
  \bibnamefont{{Schneider}}}, \bibinfo{journal}{\prd}
  \textbf{\bibinfo{volume}{74}}, \bibinfo{pages}{063520}
  (\bibinfo{year}{2006}), \eprint{arXiv:astro-ph/0607572}.

\bibitem[{\citenamefont{{Pietrobon} et~al.}(2006)\citenamefont{{Pietrobon},
  {Balbi}, and {Marinucci}}}]{piet06}
\bibinfo{author}{\bibfnamefont{D.}~\bibnamefont{{Pietrobon}}},
  \bibinfo{author}{\bibfnamefont{A.}~\bibnamefont{{Balbi}}}, \bibnamefont{and}
  \bibinfo{author}{\bibfnamefont{D.}~\bibnamefont{{Marinucci}}},
  \bibinfo{journal}{\prd} \textbf{\bibinfo{volume}{74}},
  \bibinfo{pages}{043524} (\bibinfo{year}{2006}),
  \eprint{arXiv:astro-ph/0606475}.

\bibitem[{\citenamefont{{Giannantonio}
  et~al.}(2008)\citenamefont{{Giannantonio}, {Scranton}, {Crittenden},
  {Nichol}, {Boughn}, {Myers}, and {Richards}}}]{gian08}
\bibinfo{author}{\bibfnamefont{T.}~\bibnamefont{{Giannantonio}}},
  \bibinfo{author}{\bibfnamefont{R.}~\bibnamefont{{Scranton}}},
  \bibinfo{author}{\bibfnamefont{R.~G.} \bibnamefont{{Crittenden}}},
  \bibinfo{author}{\bibfnamefont{R.~C.} \bibnamefont{{Nichol}}},
  \bibinfo{author}{\bibfnamefont{S.~P.} \bibnamefont{{Boughn}}},
  \bibinfo{author}{\bibfnamefont{A.~D.} \bibnamefont{{Myers}}},
  \bibnamefont{and} \bibinfo{author}{\bibfnamefont{G.~T.}
  \bibnamefont{{Richards}}}, \bibinfo{journal}{\prd}
  \textbf{\bibinfo{volume}{77}}, \bibinfo{pages}{123520}
  (\bibinfo{year}{2008}), \eprint{0801.4380}.

\bibitem[{\citenamefont{{Cabr{\'e}} et~al.}(2006)\citenamefont{{Cabr{\'e}},
  {Gazta{\~n}aga}, {Manera}, {Fosalba}, and {Castander}}}]{cabr06}
\bibinfo{author}{\bibfnamefont{A.}~\bibnamefont{{Cabr{\'e}}}},
  \bibinfo{author}{\bibfnamefont{E.}~\bibnamefont{{Gazta{\~n}aga}}},
  \bibinfo{author}{\bibfnamefont{M.}~\bibnamefont{{Manera}}},
  \bibinfo{author}{\bibfnamefont{P.}~\bibnamefont{{Fosalba}}},
  \bibnamefont{and}
  \bibinfo{author}{\bibfnamefont{F.}~\bibnamefont{{Castander}}},
  \bibinfo{journal}{\mnras} \textbf{\bibinfo{volume}{372}},
  \bibinfo{pages}{L23} (\bibinfo{year}{2006}), \eprint{arXiv:astro-ph/0603690}.

\bibitem[{\citenamefont{{Kaiser}}(1992)}]{kaiser92}
\bibinfo{author}{\bibfnamefont{N.}~\bibnamefont{{Kaiser}}},
  \bibinfo{journal}{\apj} \textbf{\bibinfo{volume}{388}}, \bibinfo{pages}{272}
  (\bibinfo{year}{1992}).

\bibitem[{\citenamefont{{Refregier}}(2003)}]{refregier03}
\bibinfo{author}{\bibfnamefont{A.}~\bibnamefont{{Refregier}}},
  \bibinfo{journal}{Ann.Rev.Astron.Astrophys.} \textbf{\bibinfo{volume}{41}},
  \bibinfo{pages}{645} (\bibinfo{year}{2003}), \eprint{arXiv:astro-ph/0307212}.

\bibitem[{\citenamefont{{Zhao} et~al.}(2010{\natexlab{b}})\citenamefont{{Zhao},
  {Giannantonio}, {Pogosian}, {Silvestri}, {Bacon}, {Koyama}, {Nichol}, and
  {Song}}}]{gbz10}
\bibinfo{author}{\bibfnamefont{G.}~\bibnamefont{{Zhao}}},
  \bibinfo{author}{\bibfnamefont{T.}~\bibnamefont{{Giannantonio}}},
  \bibinfo{author}{\bibfnamefont{L.}~\bibnamefont{{Pogosian}}},
  \bibinfo{author}{\bibfnamefont{A.}~\bibnamefont{{Silvestri}}},
  \bibinfo{author}{\bibfnamefont{D.~J.} \bibnamefont{{Bacon}}},
  \bibinfo{author}{\bibfnamefont{K.}~\bibnamefont{{Koyama}}},
  \bibinfo{author}{\bibfnamefont{R.~C.} \bibnamefont{{Nichol}}},
  \bibnamefont{and} \bibinfo{author}{\bibfnamefont{Y.}~\bibnamefont{{Song}}},
  \bibinfo{journal}{\prd} \textbf{\bibinfo{volume}{81}},
  \bibinfo{pages}{103510} (\bibinfo{year}{2010}{\natexlab{b}}),
  \eprint{1003.0001}.

\bibitem[{\citenamefont{{Diego} et~al.}(2002)\citenamefont{{Diego},
  {Mart{\'{\i}}nez-Gonz{\'a}lez}, {Sanz}, {Benitez}, and {Silk}}}]{mlim}
\bibinfo{author}{\bibfnamefont{J.~M.} \bibnamefont{{Diego}}},
  \bibinfo{author}{\bibfnamefont{E.}~\bibnamefont{{Mart{\'{\i}}nez-Gonz{\'a}lez}}},
  \bibinfo{author}{\bibfnamefont{J.~L.} \bibnamefont{{Sanz}}},
  \bibinfo{author}{\bibfnamefont{N.}~\bibnamefont{{Benitez}}},
  \bibnamefont{and} \bibinfo{author}{\bibfnamefont{J.}~\bibnamefont{{Silk}}},
  \bibinfo{journal}{\mnras} \textbf{\bibinfo{volume}{331}},
  \bibinfo{pages}{556} (\bibinfo{year}{2002}), \eprint{arXiv:astro-ph/0103512}.

\end{thebibliography}
\end{document}